%% file: multivariate-reconciliation.tex
\documentclass[a4paper,preprint,11pt,authoryear]{elsarticle}
\usepackage[margin=2cm]{geometry}
\usepackage{amsmath,longtable,float,multirow,parskip,placeins,bm,todonotes}
\usepackage[colorlinks=true, linkcolor=black, citecolor=black, urlcolor=black]{hyperref}

\DeclareMathOperator{\cov}{Cov}
\DeclareMathOperator{\Var}{Var}
\DeclareMathOperator{\vetor}{vec}
\DeclareMathOperator{\seno}{sin}
\DeclareMathOperator{\RMSSE}{RMSSE}
\DeclareMathOperator{\RMSE}{RMSE}
\DeclareMathOperator{\RelRMSE}{RelRMSE}
\DeclareMathOperator{\Base}{Base}
\DeclareMathOperator{\Rec}{Rec}
\DeclareMathOperator{\Uni}{Uni}
\allowdisplaybreaks
\journal{TBD}

\begin{document}

\begin{frontmatter}
  \title{Multivariate reconciliation for hierarchical time series}

  \author[label1,label2]{Ana Caroline Pinheiro} 
  \author[label1]{Rodrigo de Souza Bulhões}
  \author[label3]{Rob J. Hyndman}
  \author[label1,label4]{Paulo Canas Rodrigues}

  \affiliation[label1]{organization={Department of Statistics, Federal University of Bahia},
    city={Salvador},
    state={Bahia},
    country={Brazil}
  }
  \affiliation[label2]{organization={Secretary of Health of the State of Bahia},
    city={Salvador},
    state={Bahia},
    country={Brazil}
  }
  \affiliation[label3]{organization={Department of Econometrics \& Business Statistics, Monash University},
    city={Melbourne},
    state={Victoria},
    country={Australia}
  }

  \affiliation[label4]{organization={Department of Business Management, University of Pretoria},
    city={Pretoria},
    country={South Africa}
  }
  

  \begin{abstract}
    Some time series can be hierarchically organized into levels based on certain characteristics, such as geography or other attributes of interest. These series are referred to as hierarchical time series. Typically, forecasts are generated at all levels to ensure coherence, meaning that the forecasts should satisfy the same aggregation constraints as the observed data. Various approaches have been proposed to guarantee this coherence by using a set of base forecasts. The process through which these forecasts are adjusted to become coherent is known as forecast reconciliation. Similar to the univariate case, multivariate time series can also be structured hierarchically. However, all existing approaches are limited to a single variable. As a result, ensuring coherent forecasts requires reconciling each variable separately. However, this process does not account for correlations among multiple variables. To address this limitation, this paper proposes a multivariate reconciliation methodology that ensures coherent forecasts and incorporates relationships among variables. The proposed methodology was tested through numerical simulations, considering distinct scenarios within the series hierarchy and across multiple variables. Additionally, some base forecasting models were evaluated. The methodology was also applied to real employment data of admissions and dismissals in Brazil. The results demonstrated that multivariate reconciliation yielded more accurate outcomes than the other methods considered, both in simulated data and in practical applications.
  \end{abstract}

\end{frontmatter}

\newpage

\section{Introduction}

Collections of time series can often be aggregated at multiple levels based on geography, product classification, or other characteristics. For example, bicycle sales in a country can be aggregated to municipality, state, and national levels, and also by style and brand. These series are referred to as hierarchical time series, and forecasts of all series in the collection, at all levels of aggregation, are often required. In recent decades, this field of research has seen substantial growth, with significant applications in tourism, macroeconomics, energy, retail demand, supply chain, healthcare, emergencies, and accidents \citep{athanasopoulos2023forecast}.

Traditionally, forecasting hierarchical time series involved selecting a specific level of aggregation, generating forecasts at that level, and then disaggregating or aggregating the forecasts to obtain forecasts for the remaining structure. These methods are generally classified as bottom-up, top-down, or middle-out \citep{athanasopoulos2023forecast}, and are limited to using information from only one level, disregarding the inherent correlation structure in aggregation.

An alternative approach to forecasting hierarchical time series is to model all series separately, without considering the hierarchical structure. However, this has the undesirable consequence that the upper-level forecasts are not equal to the sum of the lower-level forecasts \citep{hyndman2011optimal}. For instance, the forecast of national sales may not equal the sum of the forecasts for each region or product category. The challenge, therefore, is to ensure that the forecasts are ``coherent'', meaning that they satisfy the aggregation constraints of the hierarchy: each aggregate forecast must equal the sum of its corresponding disaggregate forecasts.

A solution to this problem was proposed by \citet{hyndman2011optimal}, who introduced an approach in which independent forecasts are made for all series at all levels; these forecasts are then reconciled to ensure they are coherent.

An extension of the forecast reconciliation approach was developed by \citet{wickramasuriya2019optimal}, using a covariance matrix of forecast errors to improve the reconciled forecasts. The approach minimizes the mean squared error of the coherent forecasts across the entire collection of series. The results indicated that the proposed method performs well with simulated data and also in practical applications.

However, all the aforementioned approaches are limited to a single variable. Consider, for example, a hierarchical time series of worker admissions in a country, disaggregated by region and state/province. Reconciliation can be used to ensure coherent forecasts across all series. But what if we want to analyze both the number of admissions and the number of dismissals at the same hierarchical levels simultaneously? With current methods, forecast reconciliation is applied separately to each variable, without accounting for relationships between them.

The previously mentioned example illustrates that multivariate time series can also exhibit a hierarchical structure. However, existing models cannot incorporate relationships among multiple variables. To address this problem, the present study proposes a multivariate reconciliation method for hierarchical time series. The methodology will be tested through numerical simulations and applied to monthly data on worker admissions and dismissals in Brazil.

The rest of this paper is organized as follows. Section \ref{sec2} presents the notation used and the proposed methodology, while Section \ref{sec3} describes, step-by-step, the conduction of numerical simulations and their results. The application of the proposed methodology to the multivariate hierarchical time series of worker admissions and dismissals in Brazil is described in Section \ref{sec4}. Finally, Section \ref{sec5} presents the concluding remarks.

\section{Multivariate Reconciliation}
\label{sec2}

The following subsections present the notation used in this study and the proposed methodology for multivariate reconciliation.

\subsection{Notation}
\label{Notação}

\begin{figure}
  \centering
  \includegraphics[width=11cm]{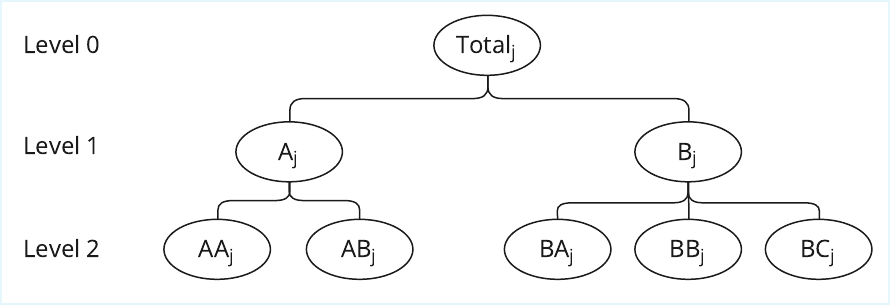}
  \caption{Multivariate hierarchical tree diagram with three levels, for $j = 1, \dots, m$.}
  \label{fig:HTS_M.ex}
\end{figure}

The notation used in this work is similar to that employed by \citet{athanasopoulos2023forecast}, but adapted for the multivariate case. Figure \ref{fig:HTS_M.ex} illustrates a multivariate hierarchical time series with three levels and with $m$ variables. For variable $j$, the $t$th observation of the total series is denoted by $y_{j,t}$, for $t = 1,\dots, T$ and $j = 1,\dots, m$. The series below the Total level is denoted by $y_{i,j,t}$, which corresponds to the $t$th observation at node $i$ for variable $j$.

It is possible to obtain, for variable $j$, the observations at time $t$
by summing the observations from the lower level. For example, for the diagram in Figure \ref{fig:HTS_M.ex}, we obtain the following equations:
\begin{align}
  y_{j,t} &= y_{A,j,t} + y_{B,j,t} \label{eq:nivel1} \\
  y_{A,j,t} & = y_{AA,j,t} + y_{AB,j,t} \label{eq:nivel2} \\
  y_{B,j,t} & = y_{BA,j,t} + y_{BB,j,t} + y_{BC,j,t} \label{eq:nivel3}
\end{align}
Substituting \eqref{eq:nivel2} and \eqref{eq:nivel3} into \eqref{eq:nivel1} we obtain
\begin{equation*}
  y_{j,t} = y_{AA,j,t} + y_{AB,j,t} + y_{BA,j,t} + y_{BB,j,t}+ y_{BC,j,t}.
\end{equation*}
These equations can also be written in the following matrix form:
\begin{equation}
  \bm{y}_{j,t} = \bm{S}\bm{b}_{j,t},
  \label{eq:ysb}
\end{equation}
where, for the variable $j$,
\begin{equation*}
  \bm{y}_{j,t} =
  \begin{bmatrix}
    y_{j,t} \\y_{A,j,t}\\y_{B,j,t}\\y_{AA,j,t}\\y_{AB,j,t}\\y_{BA,j,t}\\y_{BB,j,t}\\y_{BC,j,t}\\
  \end{bmatrix},
\end{equation*}
is an $n$-dimensional vector of all observations of variable $j$ in the hierarchy at time $t$,
\begin{equation}
  \bm{S} =
  \begin{bmatrix}
    1 & 1 & 1 & 1 & 1 \\
    1 & 1 & 0 & 0 & 0 \\
    0 & 0 & 1 & 1 & 1 \\
    1 & 0 & 0 & 0 & 0 \\
    0 & 1 & 0 & 0 & 0 \\
    0 & 0 & 1 & 0 & 0 \\
    0 & 0 & 0 & 1 & 0 \\
    0 & 0 & 0 & 0 & 1
  \end{bmatrix},
\end{equation}
is the $n\times n_b$ summation matrix, and
\begin{equation*}
  \bm{b}_{j,t} =
  \begin{bmatrix}
    y_{AA,j,t} \\y_{AB,j,t}\\y_{BA,j,t}\\y_{BB,j,t}\\y_{BC,j,t}
  \end{bmatrix},
\end{equation*}
is an $n_b$-dimensional vector of all observations of variable $j$ from the most disaggregated levels at time $t$.

The notation is applied to each variable individually. Thus, for a multivariate hierarchical time series with $m$ variables there exist $m$ bottom-level vectors $\bm{b}_{1,t},\dots,\bm{b}_{m,t}$ and $m$ all-level vectors $\bm{y}_{1,t},\dots,\bm{y}_{m,t}$. We combine the $m$ bottom-level vectors into the matrix $\bm{B}_t = [\bm{b}_{1,t},\dots,\bm{b}_{m,t}]$, with dimension $n_b \times m$, for each time $t$. Similarly, we combine the $m$ vectors $\bm{y}_{1,t},\dots,\bm{y}_{m,t}$ into the matrix $\bm{Y}_t = [\bm{y}_{1,t},\dots,\bm{y}_{m,t}]$, with dimension $n \times m$, for each time $t$. Therefore, we can express this as:
\begin{equation}
  \vetor(\bm{Y}_t) = (\bm{I}_m \otimes \bm{S})\vetor(\bm{B}_t),
  \label{eq:vec}
\end{equation}
where $\otimes$ denotes the Kronecker product, and $\vetor(\cdot)$ vectorizes the matrices by stacking their columns. Considering the hierarchy presented in Figure \ref{fig:HTS_M.ex} and $m = 2$, therefore, \eqref{eq:vec} is equivalent to:
\begin{equation*}
  \begin{bmatrix}
    y_{1,t} \\
    y_{A,1,t} \\
    y_{B,1,t} \\
    y_{AA,1,t} \\
    y_{AB,1,t} \\
    y_{BA,1,t} \\
    y_{BB,1,t} \\
    y_{BC,1,t} \\
    y_{2,t} \\
    y_{A,2,t} \\
    y_{B,2,t} \\
    y_{AA,2,t} \\
    y_{AB,2,t} \\
    y_{BA,2,t} \\
    y_{BB,2,t} \\
    y_{BC,2,t} \\
  \end{bmatrix} =
  \begin{bmatrix}
    1 & 1 & 1 & 1 & 1 & 0 & 0 & 0 & 0 & 0 \\
    1 & 1 & 0 & 0 & 0 & 0 & 0 & 0 & 0 & 0 \\
    0 & 0 & 1 & 1 & 1 & 0 & 0 & 0 & 0 & 0 \\
    1 & 0 & 0 & 0 & 0 & 0 & 0 & 0 & 0 & 0 \\
    0 & 1 & 0 & 0 & 0 & 0 & 0 & 0 & 0 & 0 \\
    0 & 0 & 1 & 0 & 0 & 0 & 0 & 0 & 0 & 0 \\
    0 & 0 & 0 & 1 & 0 & 0 & 0 & 0 & 0 & 0 \\
    0 & 0 & 0 & 0 & 1 & 0 & 0 & 0 & 0 & 0 \\
    0 & 0 & 0 & 0 & 0 & 1 & 1 & 1 & 1 & 1 \\
    0 & 0 & 0 & 0 & 0 & 1 & 1 & 0 & 0 & 0 \\
    0 & 0 & 0 & 0 & 0 & 0 & 0 & 1 & 1 & 1 \\
    0 & 0 & 0 & 0 & 0 & 1 & 0 & 0 & 0 & 0 \\
    0 & 0 & 0 & 0 & 0 & 0 & 1 & 0 & 0 & 0 \\
    0 & 0 & 0 & 0 & 0 & 0 & 0 & 1 & 0 & 0 \\
    0 & 0 & 0 & 0 & 0 & 0 & 0 & 0 & 1 & 0 \\
    0 & 0 & 0 & 0 & 0 & 0 & 0 & 0 & 0 & 1
  \end{bmatrix}
  \begin{bmatrix}
    y_{AA,1,t} \\
    y_{AB,1,t} \\
    y_{BA,1,t} \\
    y_{BB,1,t} \\
    y_{BC,1,t} \\
    y_{AA,2,t} \\
    y_{AB,2,t} \\
    y_{BA,2,t} \\
    y_{BB,2,t} \\
    y_{BC,2,t} \\
  \end{bmatrix}.
\end{equation*}
The first eight observations of $\vetor(\bm{Y}_t)$ correspond to all series in the hierarchy for variable 1, while the last eight observations correspond to variable 2. Similarly, the first five observations of $\vetor(\bm{B}_t)$ refer to the lower-level series for variable 1, while the remaining observations correspond to the lower-level series for variable 2.

\subsection{Forecast reconciliation}
\label{Reconciliação multivariada}

To obtain coherent forecasts, we first forecast all series in the hierarchy independently, without considering the hierarchical structure. These forecasts are referred to as base forecasts. These are then reconciled to make them coherent.

First, let's consider the univariate case. Let $\bm{W}_h$ be a positive definite $n \times n$ matrix, and $\widehat{\bm{y}}_{T+h|T}$ be the $h$-step-ahead base forecasts of $\bm{y}_{T+h}$ based on data up to time $T$. Linear forecast reconciliation methods for univariate time series can be expressed as \citep{wickramasuriya2019optimal}:
\begin{equation}
  \widetilde{\bm{y}}_{T+h|T} = \bm{S}(\bm{S}^\prime \bm{W}_h^{-1}\bm{S})^{-1}\bm{S}^\prime\bm{W}_h^{-1}\widehat{\bm{y}}_{T+h|T},
  \label{eq:rec_1}
\end{equation}
or equivalently, by
\begin{equation}
  \widetilde{\bm{y}}_{T+h|T} = \bm{S}[\bm{J} - \bm{J}\bm{W}_h\bm{C}^\prime(\bm{C}\bm{W}_h\bm{C}^\prime)^{-1}\bm{C}]\widehat{\bm{y}}_{T+h|T},
  \label{eq:rec_2}
\end{equation}
where the $n \times n_b$ summation matrix is partitioned as $\bm{S}^\prime =
\begin{bmatrix} \bm{A}^\prime \ \bm{I}_{n_b}
\end{bmatrix}$, $\bm{J} =
\begin{bmatrix} \bm{0}_{n_b \times n_a} \ \ \bm{I}_{n_b}
\end{bmatrix}$, $\bm{C} =
\begin{bmatrix} \bm{I}_{n_a} \ \ -\bm{A}
\end{bmatrix}$, and $n_a = n - n_b$ (the number of aggregated series).
Equation \eqref{eq:rec_2} is computationally attractive because it requires solving a system involving an $n_a \times n_a$ matrix rather than one involving the larger $n_b \times n_b$ matrix in \eqref{eq:rec_1}, and it avoids the explicit use of $\bm{W}_h^{-1}$.

It is possible to obtain another variation of \eqref{eq:rec_1} by representing the series $\bm{y}_t$ in the zero-constrained form $\bm{C}\bm{y}_t = \bm{0}_{(n_a \times 1)}$, which is equivalent to \eqref{eq:ysb}. From this, \citet{di2023cross} derived the following expression:
\begin{equation}
  \widetilde{\bm{y}}_{T+h|T} = \bm{M}_h\widehat{\bm{y}}_{T+h|T}, \quad \bm{M}_h = \bm{I}_n - \bm{W}_h\bm{C}^\prime(\bm{C}\bm{W}_h\bm{C}^\prime)^{-1}\bm{C}.
  \label{eq:rec_3}
\end{equation}
Just like \eqref{eq:rec_2}, \eqref{eq:rec_3} will have better computational performance compared to \eqref{eq:rec_1}, as it also requires the computation of only one matrix inversion, in this case, an $n_a \times n_a$ matrix.

When $\bm{W}_h$ is the covariance matrix of the $h$-step-ahead base forecast errors, the resulting forecasts are optimal in the sense that the sum of the variances of the reconciled forecasts is minimized \citep{wickramasuriya2019optimal}. This method is known as MinT (minimum trace) reconciliation.

Now we will extend the reconciliation process to the multivariate case. Let $\bm{S}_* = (\bm{I}_m \otimes \bm{S})$, where $\bm{S}$ is the summing matrix that represents the hierarchical structure, and $\bm{I}_m$ is an identity matrix of order $m$. Let $\bm{C}_* = \bm{I}_m \otimes \bm{C}$ and $\bm{J}_* = \bm{I}_m \otimes \bm{J}$. The multivariate coherence constraint can then be written as $\bm{C}_*\vetor(\bm{Y}_t) = \bm{0}_{(mn_a \times 1)}$. Let $\widehat{\bm{Y}}_{T+h|T}$ denote the $h$-step-ahead base forecasts of $\bm{Y}_{T+h}$, based on data up to time $T$. Then, the multivariate reconciled forecast can be written in one of the following three equivalent ways:
\begin{align}
  \vetor(\widetilde{\bm{Y}}_{T+h|T}) &= \bm{S}_* [\bm{S}_*^\prime \bm{W}_h^{-1} \bm{S}_*]^{-1} \bm{S}_*^\prime \bm{W}_h^{-1} \vetor(\widehat{\bm{Y}}_{T+h|T});
  \label{eq:rec_MV} \\
  &= \bm{S}_* [\bm{J}_* - \bm{N}_h] \vetor(\widehat{\bm{Y}}_{T+h|T}),
  \label{eq:rec_MV2} \\
  & = \bm{M}_h^* \vetor(\widehat{\bm{Y}}_{T+h|T}),
  \label{eq:rec_MV3}
\end{align}
where
\[
\bm{N}_h = \bm{J}_*\bm{W}_h\bm{C}_*^\prime(\bm{C}_*\bm{W}_h\bm{C}_*^\prime)^{-1}\bm{C}_*,
\]
\[
\bm{M}_h^* = \bm{I}_{nm} - \bm{W}_h\bm{C}_*^\prime(\bm{C}_*\bm{W}_h\bm{C}_*^\prime)^{-1}\bm{C}_*,
\]
and the $h$-step base forecast error vector is
\[
\bm{e}_{t,h}=\vetor(\bm{Y}_t)-\vetor(\widehat{\bm{Y}}_{t|t-h}), \qquad \bm{W}_h=\Var(\bm{e}_{t,h}).
\]
Thus, $\bm{W}_h$ is the covariance matrix of the $h$-step base forecast errors, structured in the same way as the vectorized data.

To estimate $\bm{W}_h$ we can use the same two approaches that are employed in the univariate case \citep{wickramasuriya2019optimal}:
\begin{itemize}
  \item\textbf{Sample covariance:} Let $\bm{W}_h = k_h\hat{\bm{W}}_1$ for all $h$, where $k_h > 0$ and $\widehat{\bm{W}}_1$ is the sample covariance matrix of the one-step base forecast errors. In this approach, the covariance matrices of the errors are considered proportional to each other.
  \item\textbf{Shrinkage estimator:} Let $\bm{W}_h = k_h\hat{\bm{W}}_{1,D}^*$, for all $h$, where $k_h > 0$, $\hat{\bm{W}}_{1,D}^* = \lambda_D\hat{\bm{W}}_{1,D}+(1-\lambda_D)\hat{\bm{W}}_{1}$ is the shrinkage estimator, $\hat{\bm{W}}_{1,D}$ is a diagonal matrix composed of the diagonal of $\hat{\bm{W}}_{1}$, and $\lambda_D$ is the shrinkage intensity parameter \citep{wickramasuriya2019optimal}. This estimator aims to shrink the sample covariance towards a diagonal matrix. The quantity $\lambda_D$ is estimated using the sample correlation of the residuals \citep{schafer2005shrinkage}, being computed as follows:
    \begin{equation*}
      \hat{\lambda}_D = \frac{\sum_{i\ne j}\widehat{\Var}(\hat{r}_{ij})}{\sum_{i\ne j}\hat{r}_{ij}^2},
    \end{equation*}
    where $\hat{r}_{ij}$ is the $(i,j)$th entry of $\hat{\bm{R}}_1$, which represents the one-step-ahead sample correlation matrix. When necessary, $\hat{\lambda}_D$ is restricted to the interval $[0,1]$.
\end{itemize}

In the multivariate case, the matrix $\bm{W}_h$ contains the relationships among all series in the hierarchy, including the $m$ variables. Thus, reconciliation using \eqref{eq:rec_MV} and the two approaches described previously is referred to as multivariate reconciliation in this paper.

Forecast reconciliation can be interpreted as optimally combining the base forecasts under the coherence constraint \citep{hyndman2011optimal}. In the multivariate case \eqref{eq:rec_MV}--\eqref{eq:rec_MV3}, the reconciliation process involves all series across all nodes, and so it can be seen as an optimal combination of forecasts across all series and variables in the hierarchy, accounting for relationships between and within nodes.

\section{Numerical Simulation}
\label{sec3}
To evaluate the performance of multivariate reconciliation, numerical simulations were conducted across nine scenarios. The proposed multivariate reconciliation methodology was compared to base forecasts and univariate reconciliation. Section \ref{sec3.1} outlines the step-by-step process for simulating a multivariate hierarchical time series and describes the scenarios considered in this study. Section \ref{sec3.2} summarizes the models used to obtain base forecasts for the series in the hierarchy. The evaluation metrics are presented in Section \ref{sec3.3}. Finally, the results are discussed in Section \ref{sec3.4}.

\subsection{Simulation}
\label{sec3.1}
Generating a multivariate hierarchical time series first requires simulating the lower-level series that are correlated over time, across nodes, and among multivariate observations. A Vector Autoregressive (VAR) model can be used for this purpose, with the error term incorporating relevant cross-node correlations via a Kronecker-product-based covariance matrix. What follows describes a VAR(1) simulation of $m$ series across $n_b$ bottom-level nodes.

Let the series at node $i$ be denoted $\bm{b}_{i,t}$, and assume that it follows a sinusoidal model with VAR(1) errors:
\begin{equation}
  \begin{aligned}
    \bm{b}_{i,t} & = \alpha_i \seno\bigg(\frac{2\pi t}{p}\bigg)\bm{1}_m + \bm{\eta}_{i,t}, \\
    \bm{\eta}_{i,t} & = \bm{\Phi} \bm{\eta}_{i,t-1} + \bm{\varepsilon}_{i,t},
  \end{aligned}
  \label{eq:b_it}
\end{equation}
where $\bm{1}_m$ denotes an $m$-dimensional vector of ones, $\alpha_i \sim \text{Uniform}(0,4)$ is drawn independently for each node $i$, $\bm{\varepsilon}_{i,t}$ is multivariate Gaussian white noise, $p$ represents the seasonal period, and $\bm{\Phi}$ is an $m \times m$ matrix of autoregressive coefficients. By combining the transposed noise vectors $\bm{\varepsilon}_{1,t}^\prime, \dots, \bm{\varepsilon}_{n_b,t}^\prime$ to form the rows of the matrix $\bm{E}_t$, we obtain a noise matrix of dimension $n_b \times m$. Equivalently, $\bm{\varepsilon}_{i,t}=\bm{E}_{t,i\cdot}^\prime$, where $\bm{E}_{t,i\cdot}$ is the $i$th row of $\bm{E}_t$. The covariance matrix of $\vetor(\bm{E}_t)$ can be written as:
\begin{equation*}
  \bm{W} = \cov(\vetor(\bm{E}_t)) = \bm{V} \otimes \bm{\Sigma},
\end{equation*}
where $\bm{V}$ is an $m \times m$ covariance matrix that contains the correlation structure between series at each node, and $\bm{\Sigma}$ is an $n_b \times n_b$ covariance matrix that represents the nodal correlation structure between nodes. Since $\vetor(\cdot)$ stacks columns, this implies that $\cov(E_{i,j,t},E_{k,\ell,t})=\Sigma_{ik}V_{j\ell}$.
Then, proceed as follows to simulate a realization of this structure:
\begin{enumerate}
  \item Generate or specify the covariance matrices: an $m \times m$ matrix $\bm{V}$ and an $n_b \times n_b$ matrix $\bm{\Sigma}$;
  \item Simulate $T+H$ random vectors, each of length $n_b m$, from $N(\bm{0}, \bm{V} \otimes \bm{\Sigma})$;
  \item Unstack these vectors into matrices $\bm{E}_1, \dots, \bm{E}_{T+H}$, each of size $n_b \times m$;
  \item Simulate $\bm{b}_{i,t}$ using \eqref{eq:b_it}, where $\bm{\varepsilon}_{i,t}=\bm{E}_{t,i\cdot}^\prime$, $i=1,\dots,n_b$;
  \item Combine the $m$-vectors $\bm{b}_{i,t}$ in a matrix $\bm{B}_t$, of size $n_b \times m$, and apply $\vetor(\bm{B}_t)$;
  \item Apply \eqref{eq:vec} to create $\bm{Y}_t$ using the $(\bm{I}_m \otimes \bm{S})\vetor(\bm{B}_t)$ operation.
\end{enumerate}

We conducted a simulation for a bivariate series ($m = 2$) with the same hierarchical structure as shown in Figure \ref{fig:HTS_M.ex}. Different structures were tested for the matrices $\bm{V}$ and $\bm{\Sigma}$. For $\bm{V}$, we can consider the following three cases: there is no correlation between the series at each node; there exists a positive correlation between series; there exists a negative correlation between series. For $\bm{\Sigma}$, we can consider the following three cases: there is no correlation between nodes; there is a positive correlation between nodes connected in the hierarchy; or there is a negative correlation between nodes connected in the hierarchy. Considering these structures, we specify the following matrices:
\begin{align}
  \bm{V}_1 & =
  \begin{bmatrix}
    1 & 0 \\
    0 & 1
  \end{bmatrix}, \label{eq:V1} \\
  \bm{V}_2 & =
  \begin{bmatrix}
    1 & 0.7 \\
    0.7 & 1
  \end{bmatrix}, \label{eq:V2} \\
  \bm{V}_3 & =
  \begin{bmatrix}
    1 & -0.7 \\
    -0.7 & 1
  \end{bmatrix}, \label{eq:V3} \\
  \bm{\Sigma}_1 & =
  \begin{bmatrix}
    1 & 0 & 0 & 0 & 0 \\
    0 & 1 & 0 & 0 & 0 \\
    0 & 0 & 1 & 0 & 0 \\
    0 & 0 & 0 & 1 & 0 \\
    0 & 0 & 0 & 0 & 1
  \end{bmatrix}, \label{eq:Sigma1} \\
  \bm{\Sigma}_2 & =
  \begin{bmatrix}
    1 & 0.7 & 0 & 0 & 0 \\
    0.7 & 1 & 0 & 0 & 0 \\
    0 & 0 & 1 & 0.7 & 0.7 \\
    0 & 0 & 0.7 & 1 & 0.7 \\
    0 & 0 & 0.7 & 0.7 & 1
  \end{bmatrix}, \label{eq:Sigma2} \\
  \bm{\Sigma}_3 & =
  \begin{bmatrix}
    1 & -0.4 & 0 & 0 & 0 \\
    -0.4 & 1 & 0 & 0 & 0 \\
    0 & 0 & 1 & -0.4 & -0.4 \\
    0 & 0 & -0.4 & 1 & -0.4 \\
    0 & 0 & -0.4 & -0.4 & 1
  \end{bmatrix}. \label{eq:Sigma3}
\end{align}
Thus, there are nine possible combinations of $\bm{V}$ and $\bm{\Sigma}$. Table \ref{tab:v_sigma} presents the nine scenarios used in the simulation, corresponding to these combinations.

\begin{table}[!hbt]
  \centering
  \begin{tabular}{cc}
    \hline
    \textbf{Scenario} & \textbf{Combination} \\
    \hline
    1 & $\bm{V}_1$ and $\bm{\Sigma}_1$ \\
    2 & $\bm{V}_1$ and $\bm{\Sigma}_2$ \\
    3 & $\bm{V}_1$ and $\bm{\Sigma}_3$ \\
    4 & $\bm{V}_2$ and $\bm{\Sigma}_1$ \\
    5 & $\bm{V}_2$ and $\bm{\Sigma}_2$ \\
    6 & $\bm{V}_2$ and $\bm{\Sigma}_3$ \\
    7 & $\bm{V}_3$ and $\bm{\Sigma}_1$ \\
    8 & $\bm{V}_3$ and $\bm{\Sigma}_2$ \\
    9 & $\bm{V}_3$ and $\bm{\Sigma}_3$ \\
    \hline
  \end{tabular}
  \caption{Possible combinations of $\bm{V}$ and $\bm{\Sigma}$.}
  \label{tab:v_sigma}
\end{table}

For each scenario in Table \ref{tab:v_sigma}, 1,000 simulations were conducted with total sample length $T+H = 120$ and a period of $p = 4$, with forecasts computed from the first $T$ observations for 1 to $H=12$ steps ahead. For $\bm{\Phi}$ we considered the following matrix:
\begin{equation*}
  \bm{\Phi} =
  \begin{bmatrix}
    0.7 & 0.2 \\
    0.2 & 0.7 \\
  \end{bmatrix}.
\end{equation*}

\subsection{Base forecasts}
\label{sec3.2}

To conduct multivariate reconciliation, it is necessary to obtain base forecasts for each simulated series. For this purpose, two univariate time series models were considered: the Autoregressive Integrated Moving Average (ARIMA) model and the Exponential Smoothing (ETS) state space class of models \citep{hyndman2008automatic,hyndman2021forecasting}. Additionally, since we are working with multivariate time series, a VAR model \citep{sims1980macroeconomics} was also included in the analysis.

All forecasts were computed using the \texttt{fable} package in {\sf R}. ARIMA, ETS, and VAR models were automatically selected via the \texttt{ARIMA}, \texttt{ETS}, and \texttt{VAR} functions, respectively, each with their default settings.

\subsection{Evaluation metrics}
\label{sec3.3}

The simulated dataset of length 120 was split into training and test sets, with the first $T=108$ observations used for training and the last 12 for testing. For each series, base forecasts from 1 to 12 steps ahead were generated using the models described in Section \ref{sec3.2}.

We are interested in evaluating whether the forecasts obtained through multivariate reconciliation outperform the base forecasts, disregarding any hierarchical structure. Additionally, we want to assess whether multivariate reconciliation outperforms univariate reconciliation when applied to each variable separately.

To compare multivariate reconciliation with base forecasts, we can use the following skill score \citep{hyndman2021forecasting}:
\begin{equation}
  \RelRMSE^{\Base}_{i,j,h} = 1 - \frac{\RMSE^{\Rec}_{i,j,h}}{\RMSE^{\Base}_{i,j,h}},
  \label{eq:RelRMSE_base}
\end{equation}
where $\RMSE^{\Rec}_{i,j,h}$ is the Root Mean Squared Error (RMSE) of forecasts obtained with multivariate reconciliation for node $i$ and variable $j$ at horizon $h = 1, \dots, 12$, and $\RMSE^{\Base}_{i,j, h}$ is the $\RMSE$ of the corresponding base forecasts. In the simulation study, for $R=1000$ replications,
\begin{equation*}
\RMSE^{\Rec}_{i,j,h}=\sqrt{\frac{1}{R}\sum_{r=1}^{R}\left(y^{(r)}_{i,j,T+h}-\widetilde{y}^{(r)}_{i,j,T+h|T}\right)^2},
\end{equation*}
with analogous definitions for the base and univariate reconciled forecasts. A $\RelRMSE^{\Base}$ value less than 0 indicates that the base forecasts were more accurate than those obtained through multivariate reconciliation.

Similar to \eqref{eq:RelRMSE_base}, we can compare multivariate reconciliation with univariate reconciliation using the following skill score:
\begin{equation}
  \RelRMSE^{\Uni}_{i,j,h} = 1 - \frac{\RMSE^{\Rec}_{i,j,h}}{\RMSE^{\Uni}_{i,j,h}},
  \label{eq:RelRMSE_uni}
\end{equation}
where $\RMSE^{\Uni}_{i,j,h}$ represents the $\RMSE$ of forecasts obtained with univariate reconciliation for node $i$ and variable $j$ at horizon $h$. A $\RelRMSE^{\Uni}$ value less than 0 indicates that univariate reconciliation outperformed multivariate reconciliation.

To assess which base forecasting model generates the most accurate reconciled forecasts, a scale-independent metric must be used, as the series within the hierarchy exhibit different magnitudes. One suitable option is the Root Mean Squared Scaled Error (RMSSE), which enables comparisons across series with different scales \citep{hyndman2021forecasting}. For seasonal data, the RMSSE for horizon $h$ is defined by:
\begin{equation*}
  \RMSSE_h = \sqrt{\frac{1}{mn}\sum_{i=1}^n\sum_{j=1}^{m}q^2_{i,j,h}},
\end{equation*}
where
\begin{equation*}
  q^2_{i,j,h} = \frac{(y_{i,j,T+h} - \widetilde{y}_{i,j,T+h|T})^2}{\frac{1}{T - p}\sum_{t = p+1}^{T}(y_{i,j,t} - y_{i,j,t-p})^2},
\end{equation*}
for $i=1,\dots,n$ and $j=1,\dots,m$,  where the denominator represents the mean of the squared differences in the training data for node $i$ and variable $j$, with a seasonal period of $p$. In the simulation results, these RMSSE values are averaged over the replications.

\begin{figure}
  \centering
  \includegraphics[width=\textwidth]{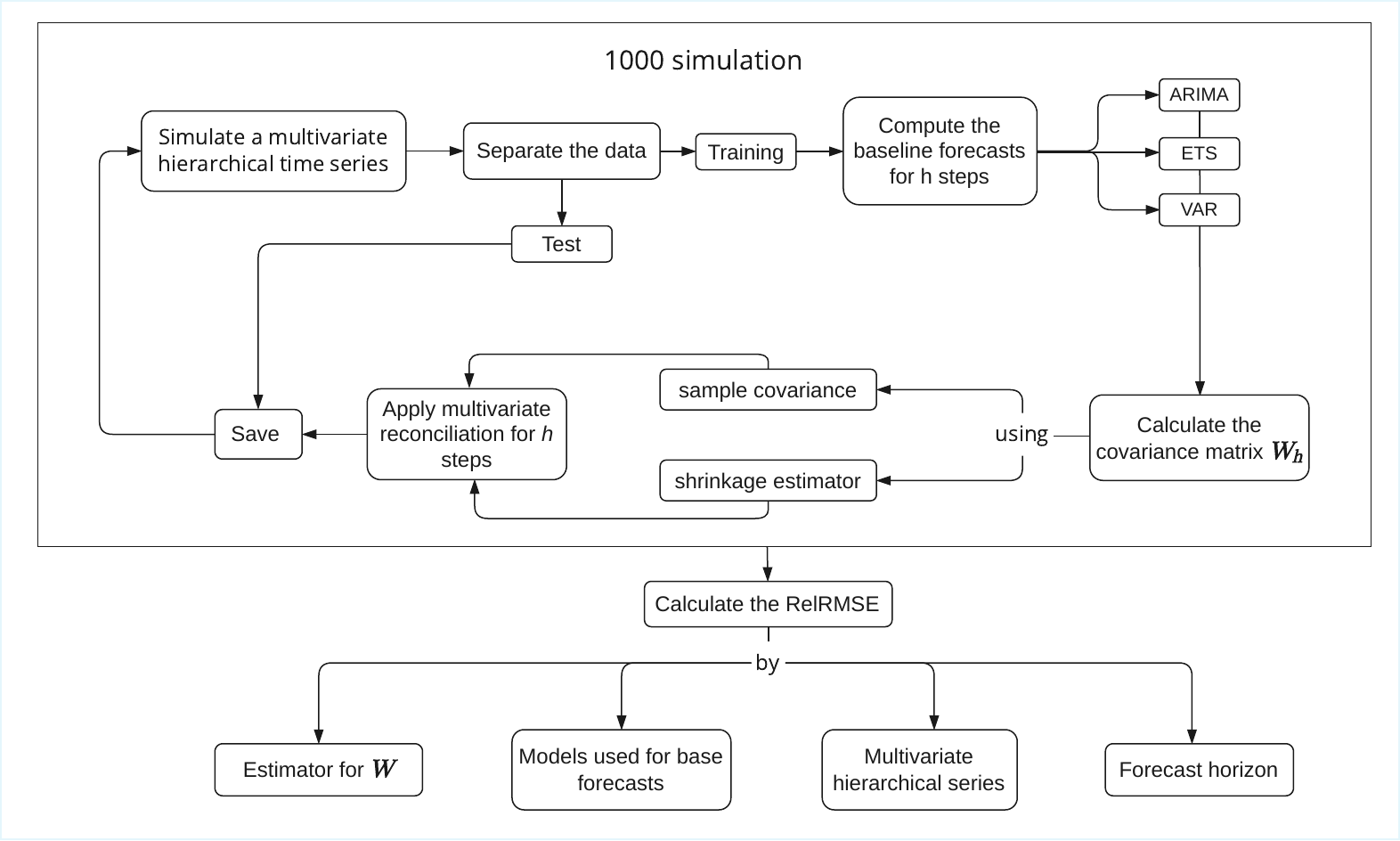}
  \caption{Flowchart of the numerical simulation.}
  \label{fig:diagrama_sim}
\end{figure}

Figure \ref{fig:diagrama_sim} presents the detailed flowchart of the numerical simulation adopted for the nine scenarios in Table \ref{tab:v_sigma}.

\subsection{Results}
\label{sec3.4}

This section presents the results of the numerical simulations, conducted according to the following steps:
\begin{enumerate}
  \item One of the nine scenarios is selected, thus defining the matrices $\bm{V}$ and $\bm{\Sigma}$;
  \item A hierarchical multivariate time series is simulated with the same structure as Figure \ref{fig:HTS_M.ex}, and with $m = 2$;
  \item The resulting 16 series are divided into training and test sets;
  \item Base forecasts from 1 to 12 steps ahead are generated for each series using the ARIMA, ETS, and VAR models;
  \item For each set of forecasts, multivariate reconciliation is applied, using both the sample covariance and the shrinkage estimator to estimate $\bm{W}_h$;
  \item For comparison purposes, univariate reconciliation is also applied using the same estimators.
\end{enumerate}
These steps are repeated 1,000 times for each scenario presented in Table \ref{tab:v_sigma}.

Figures S1 to S9 of the supplementary material present examples of the simulation of a hierarchical multivariate time series, with the same structure as Figure \ref{fig:HTS_M.ex} and with $m=2$, under scenarios 1 to 9, respectively.

Tables S1 to S9 of the supplementary material present the $\RelRMSE^{\Base}$ for all 16 hierarchical series across each forecast horizon, using ARIMA, ETS, and VAR models for base forecasts and the shrinkage approach to estimate $\bm{W}_h$ in scenarios 1 to 9, respectively. Values highlighted in red indicate a $\RelRMSE^{\Base}$ less than 0, meaning that the base forecasts achieved a lower RMSE than forecasts obtained through multivariate reconciliation.

\input{Tabelas/Tabs_sh_mean}

\input{Tabelas/Tabs_porc_sh}

Another summary of Tables S1 to S9 can be made by considering the percentage of non-negative $\RelRMSE^{\Base}$ values for each of the nine scenarios and base forecasting models (Table \ref{tab:porc_sh}). The ARIMA model exhibited 100\% of values greater than or equal to 0 in four scenarios: 2, 7, 8, and 9. For the VAR, the best results were observed in scenarios 7, 8, and 9. In the ETS case, the lowest percentages occurred in scenarios 4 and 5, both of which use $\bm{V}_2$ defined in \eqref{eq:V2} with a positive correlation of 0.7 between the variables. Table \ref{tab:porc_sh} shows that all values exceeded 50\%, highlighting the advantages of multivariate reconciliation using the shrinkage approach across all three models and nine scenarios. The ARIMA model achieved the best overall results when combined with multivariate reconciliation.

Table \ref{tab:sh_mean} provides a summary of these results by presenting the average $\RelRMSE^{\Base}$ values across the 16 hierarchical series for each model, forecast horizon, and scenario. All values in the table are greater than 0, indicating that forecasts obtained through multivariate reconciliation performed better or at least as well as the base forecasts. Among the nine scenarios analyzed, scenarios 2, 5, and 8 showed the best performance with multivariate reconciliation, as indicated by the highest average $\RelRMSE^{\Base}$ values shown in Table \ref{tab:sh_mean}. These three scenarios all use $\bm{\Sigma}_2$ defined in \eqref{eq:Sigma2}, which specifies positive correlation (0.7) between nodes at the bottom level.

Among the three base forecasting models, ARIMA stood out across all scenarios and for nearly all forecast horizons, showing the highest average $\RelRMSE^{\Base}$ values. The VAR model followed, with results close to those of the ARIMA model. For the VAR, the best results were observed in scenarios 7 and 8. In contrast, although the ETS model also produced positive averages, most of these values were very close to zero, indicating that, when used as the base forecasting method, multivariate reconciliation yielded results similar to incoherent forecasts that do not account for any hierarchical structure.

We also considered the sample covariance approach for estimating $\bm{W}_h$. However, the shrinkage approach yields better results in all scenarios.

Multivariate reconciliation using the shrinkage approach was also compared with univariate reconciliation using the same estimator. Tables S10 to S18 of the supplementary material present the $\RelRMSE^{\Uni}$ for each series in the hierarchy and for each forecasting horizon in scenarios 1 to 9, comparing these methodologies.

\input{Tabelas/Tabs_sh_uni_mean}

\input{Tabelas/Tabs_porc_sh_uni}

Table \ref{tab:sh_uni_mean} provides a summary of these results by presenting the average $\RelRMSE^{\Uni}$ values across the 16 hierarchical series for each model, forecast horizon, and scenario. The ETS model did not present mean values less than 0, indicating results that were better or similar to those of Univariate Reconciliation (Table \ref{tab:sh_uni_mean}). In scenario 1, most of the values were approximately 0, indicating that the results of multivariate and univariate reconciliation were similar. This scenario uses diagonal matrices for $\bm{V}$ and $\bm{\Sigma}$, implying that there is no correlation between variables or nodes. In scenarios 4 and 5, most of the ARIMA model means were negative, whereas in scenarios 6 and 7, most were positive. The VAR model showed the weakest results relative to univariate reconciliation in scenarios 2 and 8, both of which use $\bm{\Sigma}_2$ defined in \eqref{eq:Sigma2} with positive nodal correlation; in these scenarios univariate reconciliation outperformed multivariate reconciliation for the VAR model. The highest values observed in Table \ref{tab:sh_uni_mean} correspond to the ARIMA model, indicating a more pronounced improvement of multivariate reconciliation relative to univariate reconciliation compared to the other models.

Table \ref{tab:porc_sh_uni} summarizes the results from Tables S10 to S18, presenting the percentages of values greater than or equal to 0 for each scenario and model. The ETS model achieved the highest percentages in most scenarios, except for scenarios 2 and 8. The lowest percentages for ARIMA were recorded in scenarios 4 ($\bm{V}_2$, $\bm{\Sigma}_1$; 30.2\%) and 1 ($\bm{V}_1$, $\bm{\Sigma}_1$; 43.8\%), both of which use $\bm{\Sigma}_1$ with no nodal correlation. For ETS, the lowest values were observed in scenarios 2, 5, and 8, all of which use $\bm{\Sigma}_2$ defined in \eqref{eq:Sigma2} with positive correlation between nodes. The VAR model likewise exhibited its lowest percentages in scenarios 2 and 8, also corresponding to $\bm{\Sigma}_2$.

So far, all metrics have been used to compare the proposed methodology with already established approaches. However, they do not indicate which base forecasting model produces the best multivariate reconciled forecasts. For this analysis, we consider the multivariate reconciled forecasts using the shrinkage approach for the 16 series in the hierarchy. Table \ref{tab:RMSSE} presents the average of the 16 RMSSEs per forecasting horizon for the ARIMA, ETS, and VAR models across the nine scenarios.

The lowest values for each horizon and scenario are highlighted in bold. Across all nine scenarios, the VAR model exhibited the lowest average RMSSE values for most forecasting horizons, with ETS or ARIMA occasionally achieving a lower value at the shortest horizons ($h = 1$--2). The ETS model presented the highest average RMSSE across nearly all scenarios and forecasting horizons (Table \ref{tab:RMSSE}).

The best results for multivariate reconciled forecasts were therefore obtained with the VAR model, followed by ARIMA. The lowest absolute RMSSE values were recorded in scenarios 3, 6, and 9, which all use $\bm{\Sigma}_3$ defined in \eqref{eq:Sigma3} with negative correlation between nodes, indicating that the proposed multivariate reconciliation achieved its best performance in these three cases.

\input{Tabelas/Tabs_RMSSE}

We conclude that, for this simulation study, the shrinkage approach outperforms the sample covariance. Moreover, multivariate reconciliation using the shrinkage approach yields superior results compared to those obtained solely from base forecasts, particularly for the ARIMA model, which demonstrated improvements in over 97\% of the values across all scenarios. The results also surpass or are equivalent to the forecasts generated by univariate reconciliation. Finally, the ARIMA and VAR models stood out, presenting the best reconciled forecasts.

These results provide some insight into the role of cross-series dependence in reconciliation. Since reconciliation borrows strength across series, it can partially mimic multivariate modelling, which helps explain why univariate models such as ARIMA tend to benefit more than models like VAR that already capture cross-dependencies. Regarding negatively correlated series, our results do not suggest a systematic advantage or disadvantage. For instance, scenarios 7, 8, and 9 (which use $\bm{V}_3$ with between-variable correlation $-0.7$) achieve 100\% positive $\RelRMSE^{\Base}$ values for ARIMA, which is among the best results observed. The lowest absolute RMSSE values occur in scenarios 3, 6, and 9 (those using $\bm{\Sigma}_3$ with negative nodal correlation), suggesting that the benefits of reconciliation may depend more on the nodal correlation structure than on the sign of the between-variable correlation alone.

\FloatBarrier

\section{Multivariate reconciliation: application to data on employment admissions and dismissals in Brazil}
\label{sec4}

The monthly counts of admissions and dismissals of workers in Brazil can be organized as a bivariate hierarchical time series, structured by the country's geography. Brazil is divided into five major regions, defined by the physical characteristics of its territory: North, Northeast, Midwest, Southeast, and South. In addition, the country is subdivided into 26 states and the Federal District, each of which is a federative unit.

The North region comprises seven states: Acre (AC), Amapá (AP), Amazonas (AM), Pará (PA), Rondônia (RO), Roraima (RR), and Tocantins (TO). The Northeast region includes nine states: Alagoas (AL), Bahia (BA), Ceará (CE), Maranhão (MA), Piauí (PI), Paraíba (PB), Pernambuco (PE), Rio Grande do Norte (RN), and Sergipe (SE). The Midwest region consists of the Federal District (DF) and the following states: Goiás (GO), Mato Grosso (MT), and Mato Grosso do Sul (MS). The Southeast region encompasses four states: Espírito Santo (ES), Minas Gerais (MG), Rio de Janeiro (RJ), and São Paulo (SP). Finally, the South region comprises three states: Paraná (PR), Rio Grande do Sul (RS), and Santa Catarina (SC).

\begin{figure}[!htb]
  \centering
  \includegraphics[width=14.5cm]{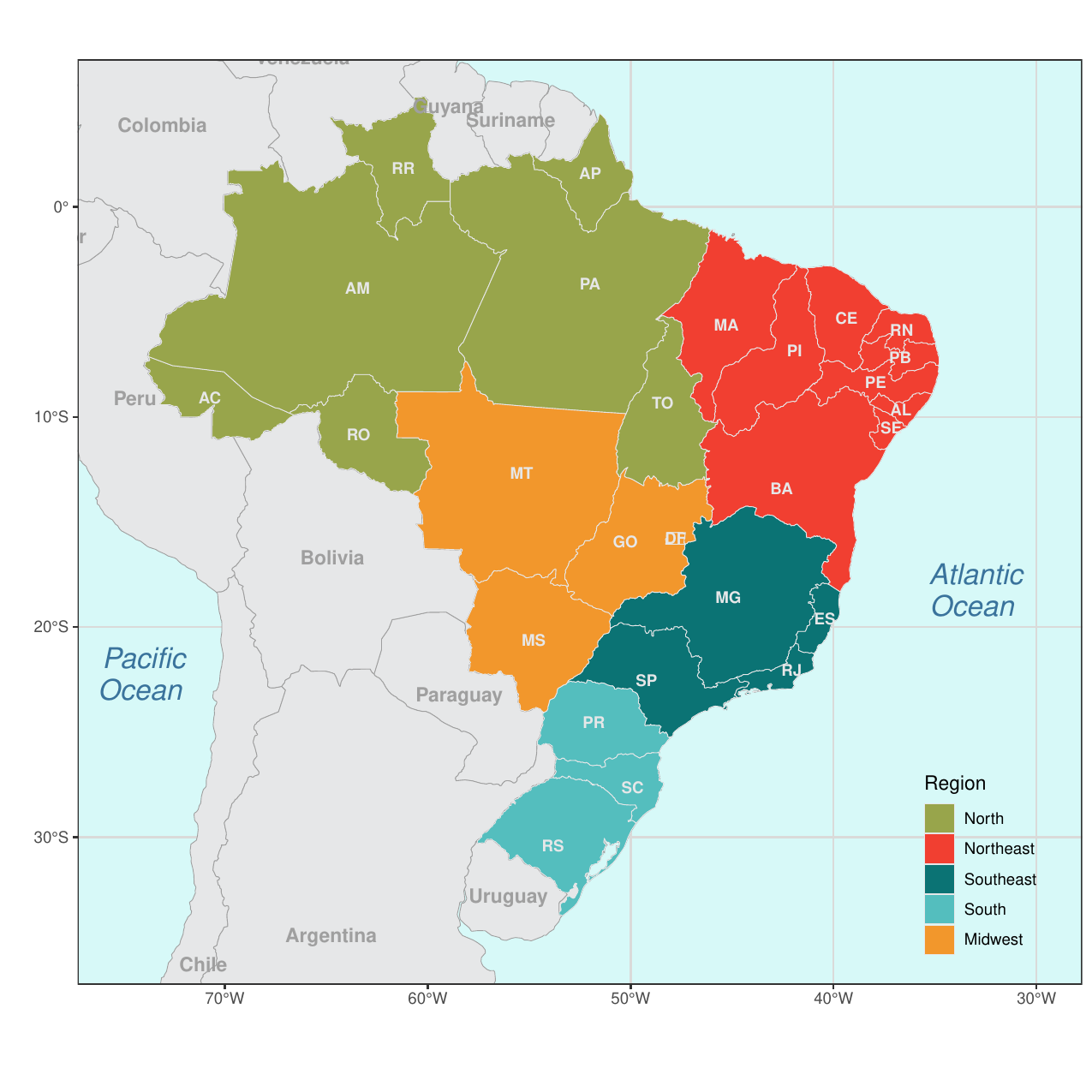}
  \caption{Location of Brazilian regions and states.}
  \label{fig:mapa_reg_uf}
\end{figure}

This structure naturally forms a three-level geographical hierarchy: Brazil, regions, and federative units. Figure \ref{fig:mapa_reg_uf} presents these geographical divisions of Brazil. Based on this structure, the admission and dismissal time series at these three levels are well-suited for evaluating the proposed method in practice. In this application, we have a bivariate hierarchical time series with three levels: at the first level, Brazil; at the second level, the five regions; and at the third level, the 27 federative units. This results in a total of 66 series: 33 admissions and 33 dismissals.

Next, Section \ref{sec4.1} presents the employment data used in the study and the process for obtaining them. Subsequently, Section \ref{sec4.2} discusses the calculation of forecast accuracy, detailing the cross-validation procedure and the evaluation metric. The exploratory data analysis is addressed in Section \ref{sec4.3}. Finally, Section \ref{sec4.4} presents the numerical results from applying the multivariate reconciliation methodology to the employment data.

\subsection{Data description}
\label{sec4.1}

Employment data are obtained through the General Register of Employed and Unemployed (CAGED) system, eSocial, and Empregador Web. CAGED is the system used by the Ministry of Labor and Employment (MTE) to record worker admissions and dismissals in Brazil. Since January 2020, the CAGED system has been partially replaced by the Digital Bookkeeping System for Tax, Social Security, and Labor Obligations (eSocial) for certain companies. Empregador Web is the system used to submit unemployment insurance claims and report dismissals to the Ministry of Labor and Employment.

To obtain the data, we used the employment database generated from CAGED declarations. This database is available through the Labor Statistics Dissemination Program (PDET). Access to the ``Bases Estatísticas RAIS e CAGED'' is provided via the Online Access tool. Through this platform, we obtained employment data (Admissions/Dismissals) by month and by federative unit from January 2004 to December 2019.

Additionally, this study also uses data from the New CAGED, which generates formal employment statistics based on information collected from the eSocial, CAGED, and Empregador Web systems. Monthly employment admission and dismissal data for Brazil, by federative unit, from January 2020 to December 2023, are available on the IpeaData platform \citep{ipeadata}.

To obtain the final database used in this study, we merged the two datasets described in the previous paragraphs. As a result, we obtained monthly employment admission and dismissal data for each federative unit from January 2004 to December 2023, totaling 20 years of data.

Through the information from the federative units, it is possible to obtain data for the five regions and the country as a whole. The total number of admissions and dismissals in Brazil throughout the year is obtained by summing the data from all federative units in each month and year. Similarly, we obtain this information by region by summing the admissions and dismissals of the federative units within each region.

\subsection{Forecast accuracy}
\label{sec4.2}

Typically, evaluating a model's forecast accuracy involves splitting the data into training and test sets, yielding a single forecast error per horizon. To obtain multiple error estimates, cross-validation was employed, enabling a more precise assessment of the proposed model. This procedure utilizes multiple training sets, each paired with a corresponding test set \citep{athanasopoulos2023forecast}.

\begin{figure}
  \centering
  \includegraphics[width=\textwidth]{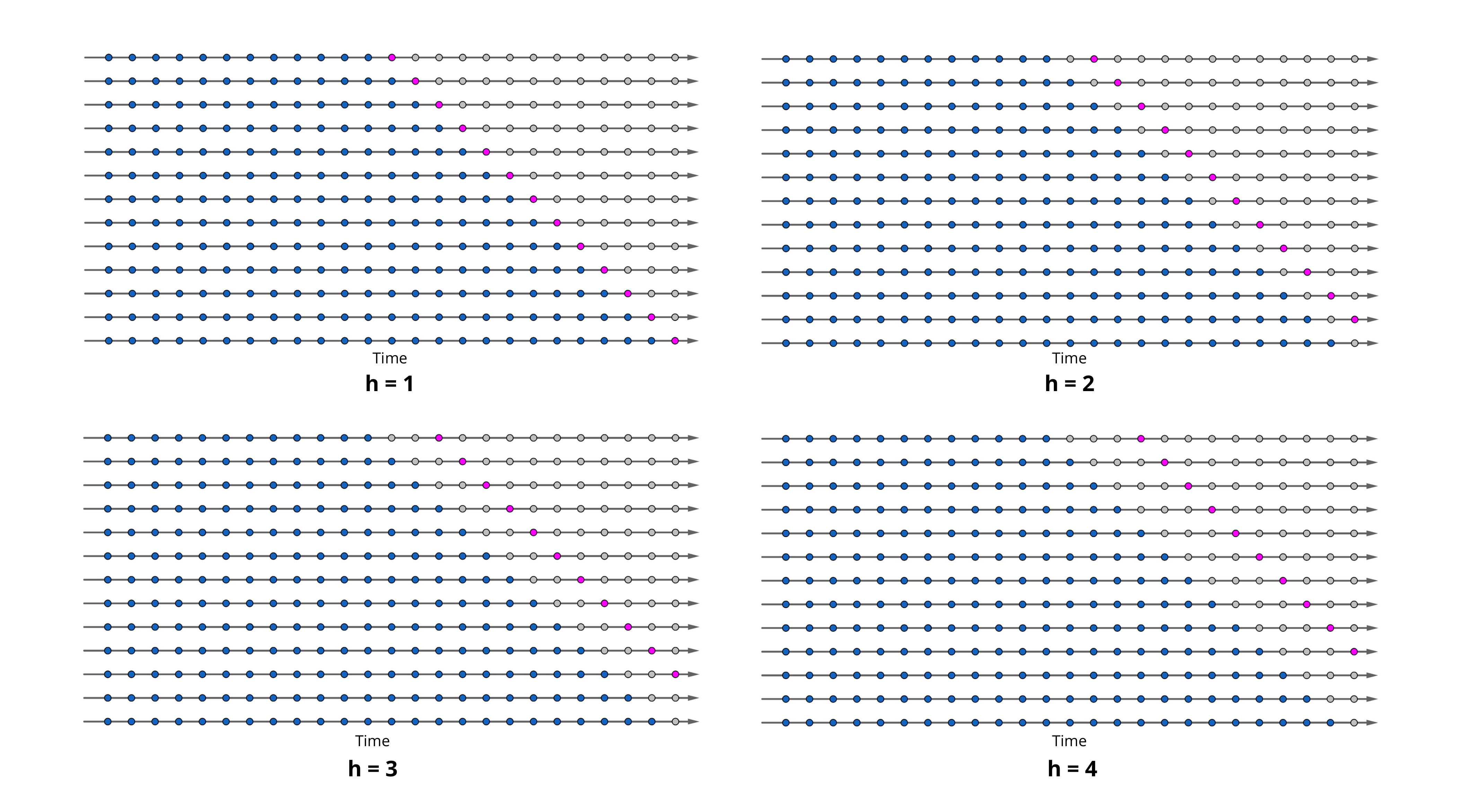}
  \caption{Procedure of cross-validation for 1 to 4 steps ahead forecasts.}
  \label{fig:validacao}
\end{figure}

Figure \ref{fig:validacao} illustrates this procedure: the blue observations represent the training sets, while the purple ones correspond to the test sets. It is noteworthy that the forecast origin shifts over time, which is why this approach is also known as ``evaluation in a rolling forecast origin''. Additionally, Figure \ref{fig:validacao} also demonstrates cross-validation applied to forecasts ranging from 1 to 4 steps ahead.

In this study, for the employment admission and dismissal dataset, cross-validation was implemented with forecasts generated for 12 origins, starting from the observation corresponding to January 2022. Forecasts were produced from 1 to 12 steps ahead. Therefore, in the empirical application, the RMSE values in \eqref{eq:RelRMSE_base} and \eqref{eq:RelRMSE_uni} are computed over the forecast origins. For example, if $t_1,\dots,t_K$ are the forecast origins, with $K=12$, then
\begin{equation*}
\RMSE^{\Rec}_{i,j,h}=\sqrt{\frac{1}{K}\sum_{k=1}^{K}\left(y_{i,j,t_k+h}-\widetilde{y}_{i,j,t_k+h|t_k}\right)^2},
\end{equation*}
with analogous definitions for base and univariate reconciled forecasts.

To evaluate the accuracy of bivariate reconciliation, we used the metrics $\RelRMSE^{\Base}$ and $\RelRMSE^{\Uni}$, defined by \eqref{eq:RelRMSE_base} and \eqref{eq:RelRMSE_uni}, respectively.

\subsection{Exploratory data analysis}
\label{sec4.3}

\begin{figure}
  \centering
  \includegraphics[width=13cm]{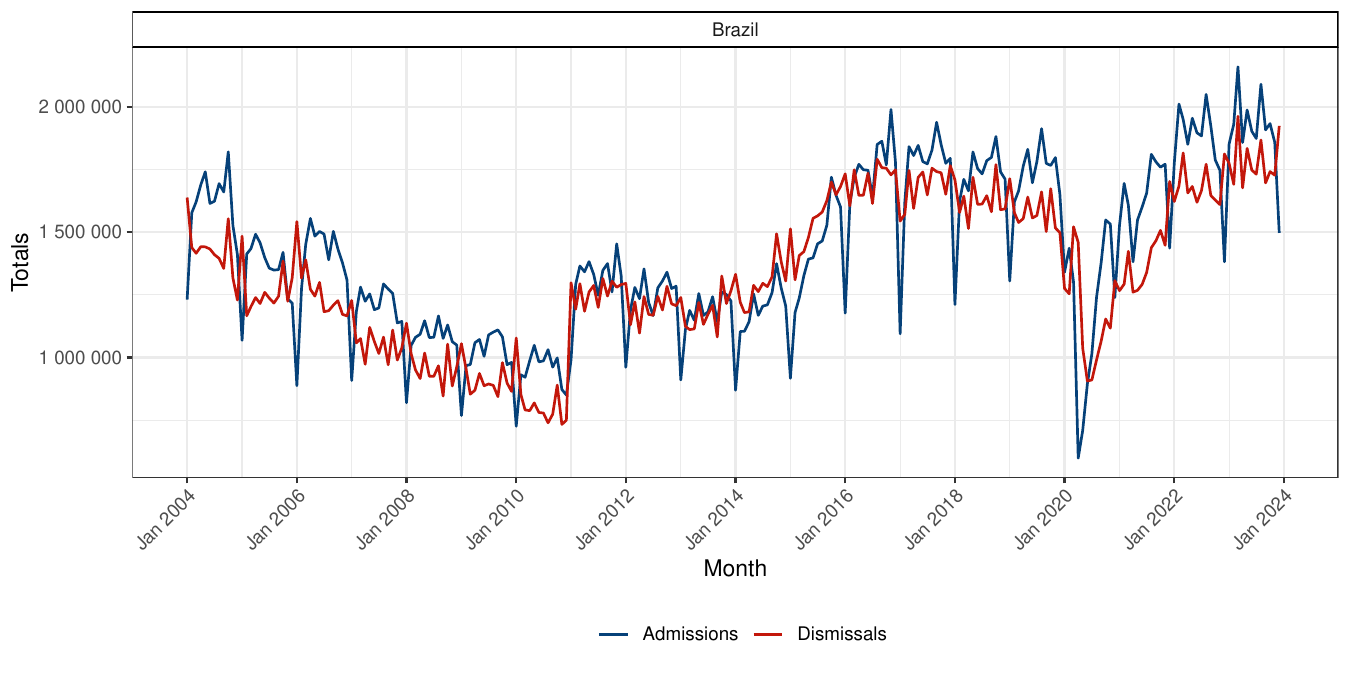}
  \caption{Time series of employment admissions and dismissals in Brazil.}
  \label{fig:serie_emprego_br}
\end{figure}

\begin{figure}
  \centering
  \includegraphics[width=\textwidth]{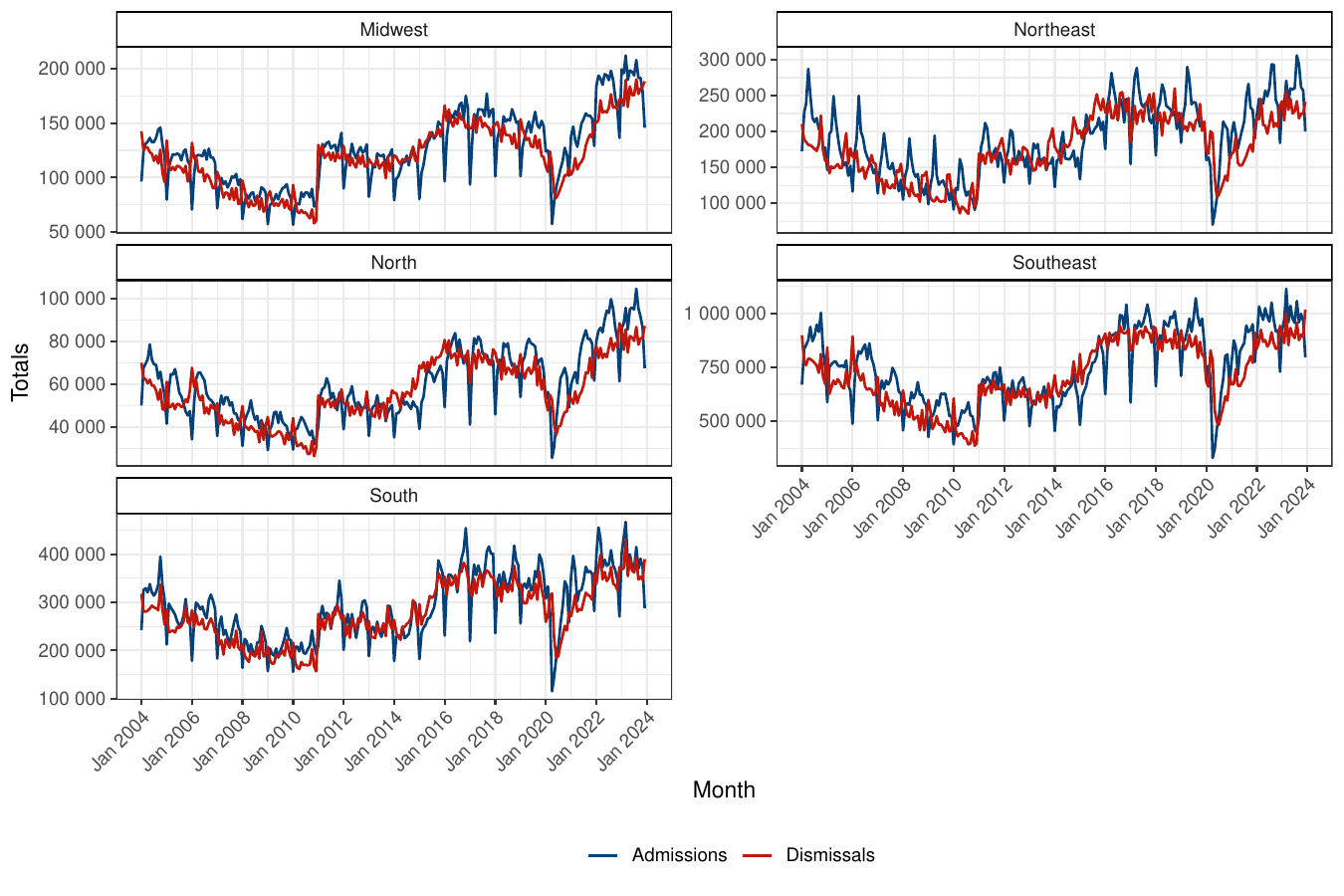}
  \caption{Time series of the numbers of employment admissions and dismissals in Brazil by region.}
  \label{fig:serie_emprego_reg}
\end{figure}

\begin{figure}[p]
  \centering
  \includegraphics[width=\textwidth]{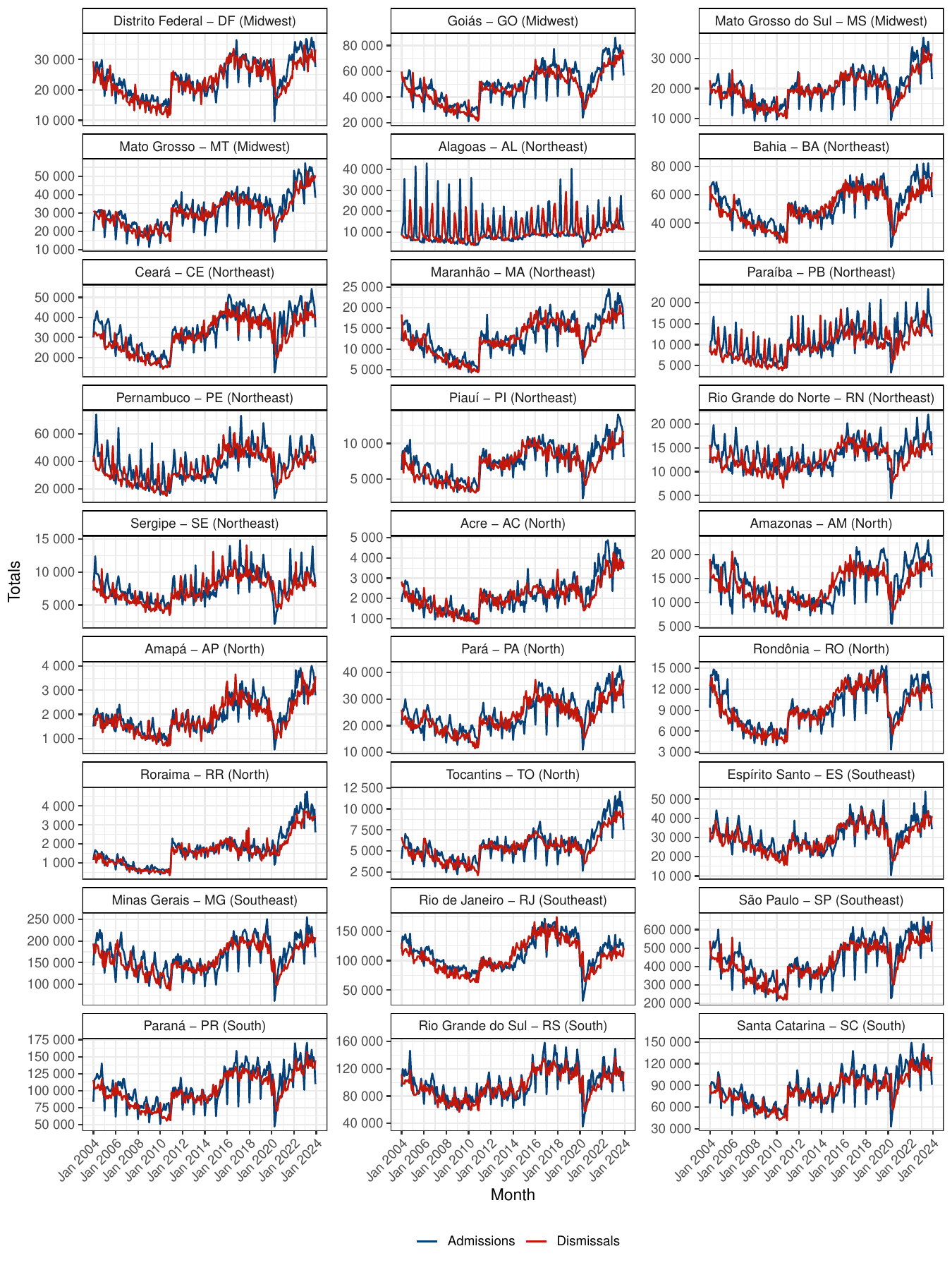}
  \caption{Time series of employment admissions and dismissals in each federative unit.}
  \label{fig:serie_emprego_uf}
\end{figure}

Figures \ref{fig:serie_emprego_br}, \ref{fig:serie_emprego_reg}, and \ref{fig:serie_emprego_uf} present the series of worker dismissals and admissions in Brazil, from January 2004 to December 2023, regions, and federative units, respectively. A decline in admissions and dismissals was observed in Brazil across regions and in almost all federative units between 2004 and 2010. These values remained stable until mid-2015, when employment figures began to increase. A sharp decline was observed in 2020, coinciding with the onset of the COVID-19 pandemic in the country (Figures \ref{fig:serie_emprego_br}, \ref{fig:serie_emprego_reg} and \ref{fig:serie_emprego_uf}). In 2023, Brazil recorded its highest number of admissions, exceeding two million, with a similar number of dismissals in the same period (Figure \ref{fig:serie_emprego_br}). Nearly all admission series exhibit a seasonal pattern that repeats each year. The admission and dismissal series follow the same trend, indicating a correlation between the series (Figures \ref{fig:serie_emprego_br}, \ref{fig:serie_emprego_reg}, and \ref{fig:serie_emprego_uf}). This phenomenon can be explained by \citet{gonzaga2014rotatividade}, who state that the Brazilian labor market presents a high job turnover, meaning frequent movement of workers between their work positions.

The admission and dismissal series for the Midwest, Northeast, North, Southeast, and South regions exhibit similar patterns, differing only in scale (Figure \ref{fig:serie_emprego_reg}). The Southeast stands out as the region with the highest numbers of admissions and dismissals. Although the Southeast comprises only four states, three of them --- São Paulo, Minas Gerais, and Rio de Janeiro --- are the most populous in Brazil, which explains the high volume of employment admissions and dismissals. Conversely, the North, being the least populated region in the country, presents the lowest values for these series.

The patterns observed in the Brazil series (Figure \ref{fig:serie_emprego_br}) and in regional series (Figure \ref{fig:serie_emprego_reg}) are also repeated in the admission and dismissal series of the federative units (Figure \ref{fig:serie_emprego_uf}). However, the series for the state of Alagoas stands out, as they exhibit a completely distinct trend and seasonality compared to the others.
Additionally, São Paulo and Minas Gerais present the highest numbers of admissions and dismissals, which can be attributed to their status as the most populous states in Brazil. Conversely, Amapá and Roraima, having smaller populations, record the lowest values for these series.

\subsection{Results}
\label{sec4.4}

Through simulation, we observed that among the three models analyzed, ARIMA, ETS, and VAR, the ARIMA and VAR models yielded the best results when applying the proposed bivariate reconciliation approach. Therefore, these models were employed to generate the base forecasts for the 66 employment time series.

Tables \ref{tab:adm_sh} and \ref{tab:dem_sh} show the results obtained using the shrinkage estimator for $\bm{W}_h$, with Table \ref{tab:adm_sh} dedicated to the admission series and Table \ref{tab:dem_sh} to the dismissal series.



\input{Tabelas/Tabs_adm_sh}

The proposed methodology, using the shrinkage estimator to estimate $\bm{W}_h$, yielded significantly better results for the admission series (Table \ref{tab:adm_sh}). In this table, 75.3\% of the values were greater than or equal to 0. Among the 33 series, 19 exhibited only one or no values less than 0 across the 12 forecast horizons. The worst results were observed in the South region and in the states of ES, PA, PE, RN, and SP. Overall, a substantial improvement was achieved by employing this methodology with the shrinkage estimator for the admission series.

\input{Tabelas/Tabs_dem_sh}

For the dismissal series, using the shrinkage estimator (Table \ref{tab:dem_sh}), the bivariate reconciliation outperformed the base forecasts in 54.8\% of the observations. The best results were observed in the series for the Midwest and South regions, as well as in the states of AL, AP, MS, MT, PB, and SC, all of which presented two or fewer values below 0. Although the results for the dismissal series were not as satisfactory as those obtained for the admission series (Table \ref{tab:adm_sh}), improvements were still observed in more than half of the cases when compared to the base forecasts.

For the employment data application, the proposed bivariate reconciliation method using the ARIMA model and the shrinkage estimator demonstrated strong results for the admission series and moderate performance for the dismissal series, highlighting its advantages over the base forecasts. These findings align with the simulation results.

The proposed methodology was also tested using the VAR model to generate the base forecasts for the admission and dismissal series. It is worth noting that the VAR model should be applied to stationary series; however, the employment series is not stationary. Thus, the VAR model was used solely for comparison purposes in this application. The results of the VAR model were placed in the supplementary material.




Tables S19 and S20 present the $\RelRMSE^{\Base}$ for the admission and dismissal variables, respectively, both using the shrinkage approach to estimate $\bm{W}_h$. It can be seen that 49.2\% of the $\RelRMSE^{\Base}$ values were non-negative for the admission variable (Table S19), indicating that the shrinkage-based methodology and the VAR model perform similarly to the base forecasts. Only the states of AL, ES, and SC showed good results across all forecast horizons with this methodology.

A similar result was observed for the dismissal variable, with 51.3\% of the values of $\RelRMSE^{\Base}$ being greater than or equal to 0 (Table S20). Additionally, the multivariate reconciliation using the VAR model and the shrinkage approach outperformed the base forecasts in all horizons in the states of AL, MS, and RO for this variable. When the VAR model is used for base forecasts, multivariate reconciliation with the shrinkage estimator produces results that are similar to those of the base forecasts.

When comparing the ARIMA and VAR models, it is observed that multivariate reconciliation combined with ARIMA produced better results. This configuration led to significant improvements in the forecasts for both the admission and dismissal variables compared to the base forecasts.


The application of multivariate reconciliation to the data of worker admissions and dismissals in Brazil demonstrates its advantage in real data, especially when used with the ARIMA model and the shrinkage approach.

\section{Conclusion}
\label{sec5}

Current methods for forecast reconciliation in hierarchical time series do not allow the incorporation of multiple variables, requiring reconciliation to be applied separately to each variable and thus neglecting the correlation among multivariate series. This study presented a multivariate reconciliation methodology capable of considering both the relationships within the hierarchy and the correlation among multiple variables.

The proposed methodology uses two estimators for the covariance matrix of base forecast errors, $\bm{W}_h$: the sample covariance and the shrinkage estimator. Additionally, three base forecasting models were evaluated: ARIMA, ETS, and VAR. To validate the approach, simulations and an application using real data were conducted.

The simulations considered nine scenarios, representing different combinations of the matrices $\bm{V}$ and $\bm{\Sigma}$, which determine the correlation between pairs of variables and between the series at the last level of the hierarchy, respectively.

Of the two estimators considered for $\bm{W}_h$, the shrinkage approach outperformed the sample covariance across all scenarios, and so proved more suitable for estimating $\bm{W}_h$ in multivariate reconciliation. When combined with multivariate reconciliation, the shrinkage estimator was shown to be effective in obtaining coherent forecasts: for all three base forecasting models (ARIMA, ETS, and VAR), this methodology outperformed the base forecasts in all nine scenarios. The ARIMA model stood out, with over 97\% of the $\RelRMSE^{\Base}$ values exceeding 0, indicating significant improvements over the base forecasts.

When comparing multivariate reconciliation with univariate reconciliation, both using the shrinkage approach, it was found that multivariate reconciliation yielded better or similar results to the univariate approach for almost all scenarios.

Among the three base forecasting models evaluated, ARIMA and VAR produced the smallest forecast errors when multivariate reconciliation was applied with the shrinkage approach.

The application of the proposed methodology to employment data in Brazil demonstrated that it can also be effective with empirical data. For these data, multivariate reconciliation with the ARIMA model and the shrinkage approach outperformed the VAR model and sample covariance-based reconciliation. Moreover, the errors of the reconciled forecasts were smaller than those of the base forecasts.

The methodology extends naturally to more than two variables, though the $(mn \times mn)$ covariance matrix $\bm{W}_h$ grows rapidly with $m$ and $n$, making scalable estimation an important direction for future work. Structured approximations (such as factor models or sparse estimators) could make the approach practical for large hierarchies. Beyond point forecasts, extending the framework to produce coherent joint predictive distributions would be a valuable next step, building on the growing literature on probabilistic forecast reconciliation. The simulation study could also be broadened to cover richer configurations of $\bm{V}$, $\bm{\Sigma}$, and $\bm{\Phi}$, as well as grouped and temporal hierarchies. These extensions would further establish multivariate reconciliation as a general-purpose tool for coherent forecasting of interconnected hierarchical systems.

\bibliographystyle{elsarticle-harv}
\bibliography{cas-refs}

\end{document}

%% file: Tabelas/Tabs_sh_mean.tex
\begin{table}[!htb]
  \caption{Average of $\RelRMSE^{\Base}$ of the 16 series for different forecast horizons and for each scenario. ARIMA, ETS, and VAR models for base forecasts. Using the shrinkage approach to estimate $\bm{W}$. All values are positive, indicating reconciliation always improves upon base forecasts, on average.}
  \centering
  \resizebox{1\textwidth}{!}{%
    \begin{tabular}{lrrrrrrrrrrrr}
      \hline
      \multirow{2}{*}{Model} & \multicolumn{12}{c}{Forecast horizons ($h$)} \\ \cline{2-13}
      & 1 & 2 & 3 & 4 & 5 & 6 & 7 & 8 & 9 & 10 & 11 & 12 \\
      \hline
      Scenario 1 & & & & & & & & & & & & \\
      \hline
      ARIMA & 0.022 & 0.018 & 0.017 & 0.013 & 0.018 & 0.017 & 0.019 & 0.017 & 0.021 & 0.018 & 0.021 & 0.020 \\
      ETS & 0.004 & 0.002 & 0.002 & 0.001 & 0.002 & 0.001 & 0.001 & 0.001 & 0.001 & 0.001 & 0.001 & 0.001 \\
      VAR & 0.013 & 0.010 & 0.011 & 0.008 & 0.013 & 0.012 & 0.015 & 0.010 & 0.016 & 0.011 & 0.013 & 0.009 \\
      \hline
      Scenario 2 & & & & & & & & & & & & \\
      \hline
      ARIMA & 0.024 & 0.023 & 0.020 & 0.014 & 0.023 & 0.022 & 0.027 & 0.026 & 0.029 & 0.032 & 0.032 & 0.032 \\
      ETS & 0.003 & 0.001 & 0.001 & 0.000 & 0.001 & 0.000 & 0.001 & 0.000 & 0.000 & 0.000 & 0.000 & -0.000 \\
      VAR & 0.017 & 0.012 & 0.013 & 0.006 & 0.016 & 0.009 & 0.014 & 0.006 & 0.017 & 0.008 & 0.017 & 0.007 \\
      \hline
      Scenario 3 & & & & & & & & & & & & \\
      \hline
      ARIMA & 0.021 & 0.020 & 0.020 & 0.014 & 0.024 & 0.018 & 0.021 & 0.017 & 0.024 & 0.021 & 0.024 & 0.021 \\
      ETS & 0.005 & 0.005 & 0.005 & 0.004 & 0.004 & 0.004 & 0.004 & 0.005 & 0.004 & 0.003 & 0.003 & 0.003 \\
      VAR & 0.006 & 0.005 & 0.005 & 0.004 & 0.007 & 0.008 & 0.007 & 0.006 & 0.007 & 0.006 & 0.006 & 0.005 \\
      \hline
      Scenario 4 & & & & & & & & & & & & \\
      \hline
      ARIMA & 0.022 & 0.015 & 0.011 & 0.012 & 0.019 & 0.018 & 0.019 & 0.019 & 0.025 & 0.024 & 0.026 & 0.023 \\
      ETS & 0.002 & 0.001 & 0.000 & 0.000 & 0.001 & 0.000 & 0.000 & 0.000 & 0.000 & -0.000 & 0.000 & 0.000 \\
      VAR & 0.011 & 0.010 & 0.008 & 0.005 & 0.011 & 0.009 & 0.009 & 0.006 & 0.010 & 0.005 & 0.007 & 0.003 \\
      \hline
      Scenario 5 & & & & & & & & & & & & \\
      \hline
      ARIMA & 0.022 & 0.020 & 0.015 & 0.011 & 0.018 & 0.020 & 0.020 & 0.020 & 0.026 & 0.028 & 0.029 & 0.029 \\
      ETS & 0.002 & 0.001 & 0.001 & 0.000 & 0.000 & 0.000 & 0.000 & -0.000 & 0.000 & -0.000 & 0.000 & 0.000 \\
      VAR & 0.021 & 0.015 & 0.010 & 0.006 & 0.014 & 0.011 & 0.013 & 0.007 & 0.014 & 0.006 & 0.013 & 0.002 \\
      \hline
      Scenario 6 & & & & & & & & & & & & \\
      \hline
      ARIMA & 0.026 & 0.020 & 0.014 & 0.013 & 0.025 & 0.021 & 0.021 & 0.020 & 0.027 & 0.023 & 0.025 & 0.019 \\
      ETS & 0.005 & 0.005 & 0.005 & 0.006 & 0.005 & 0.004 & 0.004 & 0.005 & 0.005 & 0.004 & 0.004 & 0.004 \\
      VAR & 0.004 & 0.005 & 0.003 & 0.003 & 0.004 & 0.004 & 0.003 & 0.003 & 0.004 & 0.002 & 0.002 & 0.001 \\
      \hline
      Scenario 7 & & & & & & & & & & & & \\
      \hline
      ARIMA & 0.022 & 0.019 & 0.018 & 0.015 & 0.022 & 0.021 & 0.023 & 0.020 & 0.024 & 0.023 & 0.024 & 0.023 \\
      ETS & 0.003 & 0.002 & 0.001 & 0.001 & 0.001 & 0.001 & 0.001 & 0.001 & 0.001 & 0.001 & 0.001 & 0.001 \\
      VAR & 0.016 & 0.014 & 0.016 & 0.011 & 0.020 & 0.015 & 0.020 & 0.012 & 0.023 & 0.013 & 0.020 & 0.010 \\
      \hline
      Scenario 8 & & & & & & & & & & & & \\
      \hline
      ARIMA & 0.021 & 0.017 & 0.017 & 0.014 & 0.022 & 0.023 & 0.023 & 0.021 & 0.025 & 0.026 & 0.027 & 0.025 \\
      ETS & 0.003 & 0.002 & 0.001 & 0.001 & 0.001 & 0.001 & 0.000 & 0.000 & 0.001 & 0.000 & 0.000 & 0.000 \\
      VAR & 0.025 & 0.016 & 0.020 & 0.010 & 0.025 & 0.014 & 0.022 & 0.008 & 0.026 & 0.011 & 0.024 & 0.007 \\
      \hline
      Scenario 9 & & & & & & & & & & & & \\
      \hline
      ARIMA & 0.021 & 0.020 & 0.018 & 0.013 & 0.022 & 0.016 & 0.020 & 0.017 & 0.025 & 0.019 & 0.023 & 0.019 \\
      ETS & 0.007 & 0.005 & 0.005 & 0.004 & 0.005 & 0.004 & 0.003 & 0.003 & 0.004 & 0.003 & 0.003 & 0.003 \\
      VAR & 0.008 & 0.010 & 0.010 & 0.008 & 0.010 & 0.009 & 0.009 & 0.006 & 0.010 & 0.008 & 0.009 & 0.006 \\
      \hline
    \end{tabular}%
  }
  \label{tab:sh_mean}
\end{table}

%% file: Tabelas/Tabs_porc_sh.tex
\begin{table}[!htb]
  \caption{Percentage of $\RelRMSE^{\Base}$ greater than or equal to 0 considering the 16 series in the hierarchy and the 12 forecast horizons. Using the shrinkage approach to estimate $\bm{W}$ for the 9 scenarios.}
  \centering
  \begin{tabular}{lrrrrrrrrr}
    \hline
    \multirow{2}{*}{Model} & \multicolumn{9}{c}{Scenario} \\ \cline{2-10}
    & \multicolumn{1}{c}{1} & \multicolumn{1}{c}{2} & \multicolumn{1}{c}{3} & \multicolumn{1}{c}{4} & \multicolumn{1}{c}{5} & 6 & 7 & 8 & 9 \\ \hline
    ARIMA & 97.4 & 100.0 & 97.4 & 97.4 & 99.0 & 99.0 & 100.0 & 100.0 & 100.0 \\
    ETS & 73.4 & 63.5 & 97.9 & 50.5 & 53.6 & 95.8 & 72.9 & 72.4 & 93.2 \\
    VAR & 89.1 & 77.6 & 79.7 & 63.5 & 67.7 & 55.7 & 97.9 & 92.2 & 94.8 \\
    \hline
  \end{tabular}%
  \label{tab:porc_sh}
\end{table}

%% file: Tabelas/Tabs_sh_uni_mean.tex
\begin{table}[!htb]
  \caption{Average of $\RelRMSE^{\Uni}$ of the 16 series for different forecast horizons and for each scenario. ARIMA, ETS, and VAR models for base forecasts. Using the shrinkage approach to estimate $\bm{W}$. Values in red indicate a $\RelRMSE^{\Uni}$ less than 0.}
  \centering
  \resizebox{1\textwidth}{!}{%
    \begin{tabular}{lrrrrrrrrrrrr}
      \hline
      \multirow{2}{*}{Model} & \multicolumn{12}{c}{Forecast horizons ($h$)} \\ \cline{2-13}
      & 1 & 2 & 3 & 4 & 5 & 6 & 7 & 8 & 9 & 10 & 11 & 12 \\
      \hline
      Scenario 1 & & & & & & & & & & & & \\
      \hline
      ARIMA & 0.001 & 0.001 & 0.001 & 0.000 & 0.000 & \textcolor{red}{-0.001} & \textcolor{red}{-0.000} & \textcolor{red}{-0.001} & \textcolor{red}{-0.002} & \textcolor{red}{-0.001} & \textcolor{red}{-0.001} & \textcolor{red}{-0.002} \\
      ETS & 0.001 & 0.001 & 0.001 & 0.001 & 0.000 & 0.000 & 0.000 & 0.000 & 0.000 & 0.000 & 0.000 & 0.000 \\
      VAR & 0.002 & 0.001 & \textcolor{red}{-0.000} & \textcolor{red}{-0.000} & 0.001 & \textcolor{red}{-0.000} & \textcolor{red}{-0.000} & 0.000 & 0.000 & 0.000 & 0.000 & \textcolor{red}{-0.000} \\
      \hline
      Scenario 2 & & & & & & & & & & & & \\
      \hline
      ARIMA & 0.002 & 0.001 & 0.001 & \textcolor{red}{-0.001} & 0.001 & 0.001 & 0.002 & 0.002 & 0.002 & 0.003 & 0.003 & 0.003 \\
      ETS & 0.000 & 0.000 & \textcolor{red}{-0.000} & 0.000 & \textcolor{red}{-0.000} & \textcolor{red}{-0.000} & \textcolor{red}{-0.000} & \textcolor{red}{-0.000} & \textcolor{red}{-0.000} & \textcolor{red}{-0.000} & \textcolor{red}{-0.000} & \textcolor{red}{-0.000} \\
      VAR & \textcolor{red}{-0.001} & \textcolor{red}{-0.002} & \textcolor{red}{-0.002} & \textcolor{red}{-0.002} & \textcolor{red}{-0.002} & \textcolor{red}{-0.003} & \textcolor{red}{-0.003} & \textcolor{red}{-0.003} & \textcolor{red}{-0.002} & \textcolor{red}{-0.003} & \textcolor{red}{-0.002} & \textcolor{red}{-0.002} \\
      \hline
      Scenario 3 & & & & & & & & & & & & \\
      \hline
      ARIMA & 0.002 & 0.003 & 0.003 & 0.002 & 0.002 & 0.001 & 0.001 & \textcolor{red}{-0.001} & \textcolor{red}{-0.001} & \textcolor{red}{-0.001} & \textcolor{red}{-0.002} & \textcolor{red}{-0.001} \\
      ETS & 0.001 & 0.002 & 0.003 & 0.002 & 0.002 & 0.002 & 0.002 & 0.002 & 0.001 & 0.001 & 0.001 & 0.002 \\
      VAR & \textcolor{red}{-0.002} & \textcolor{red}{-0.002} & \textcolor{red}{-0.002} & \textcolor{red}{-0.002} & \textcolor{red}{-0.002} & \textcolor{red}{-0.001} & \textcolor{red}{-0.001} & \textcolor{red}{-0.001} & \textcolor{red}{-0.001} & \textcolor{red}{-0.001} & \textcolor{red}{-0.001} & \textcolor{red}{-0.000} \\
      \hline
      Scenario 4 & & & & & & & & & & & & \\
      \hline
      ARIMA & 0.001 & \textcolor{red}{-0.002} & \textcolor{red}{-0.001} & 0.000 & \textcolor{red}{-0.001} & \textcolor{red}{-0.002} & \textcolor{red}{-0.001} & \textcolor{red}{-0.001} & \textcolor{red}{-0.002} & \textcolor{red}{-0.004} & \textcolor{red}{-0.005} & \textcolor{red}{-0.003} \\
      ETS & 0.000 & 0.000 & 0.000 & \textcolor{red}{-0.000} & 0.000 & \textcolor{red}{-0.000} & \textcolor{red}{-0.000} & 0.000 & 0.000 & \textcolor{red}{-0.000} & 0.000 & 0.000 \\
      VAR & \textcolor{red}{-0.001} & \textcolor{red}{-0.002} & \textcolor{red}{-0.001} & \textcolor{red}{-0.000} & 0.000 & \textcolor{red}{-0.000} & 0.000 & 0.000 & \textcolor{red}{-0.000} & \textcolor{red}{-0.000} & 0.001 & 0.000 \\
      \hline
      Scenario 5 & & & & & & & & & & & & \\
      \hline
      ARIMA & 0.001 & \textcolor{red}{-0.000} & \textcolor{red}{-0.001} & \textcolor{red}{-0.002} & \textcolor{red}{-0.001} & \textcolor{red}{-0.000} & \textcolor{red}{-0.001} & \textcolor{red}{-0.000} & \textcolor{red}{-0.000} & 0.001 & 0.000 & 0.001 \\
      ETS & 0.001 & 0.000 & 0.000 & 0.000 & 0.000 & 0.000 & \textcolor{red}{-0.000} & \textcolor{red}{-0.000} & \textcolor{red}{-0.000} & \textcolor{red}{-0.000} & \textcolor{red}{-0.000} & \textcolor{red}{-0.000} \\
      VAR & 0.002 & 0.000 & \textcolor{red}{-0.001} & \textcolor{red}{-0.001} & 0.001 & \textcolor{red}{-0.000} & \textcolor{red}{-0.000} & \textcolor{red}{-0.000} & 0.001 & \textcolor{red}{-0.001} & 0.000 & \textcolor{red}{-0.001} \\
      \hline
      Scenario 6 & & & & & & & & & & & & \\
      \hline
      ARIMA & 0.003 & 0.001 & \textcolor{red}{-0.000} & 0.003 & 0.005 & 0.002 & 0.001 & 0.002 & 0.001 & \textcolor{red}{-0.000} & \textcolor{red}{-0.001} & \textcolor{red}{-0.002} \\
      ETS & 0.002 & 0.002 & 0.003 & 0.003 & 0.003 & 0.002 & 0.002 & 0.002 & 0.002 & 0.002 & 0.002 & 0.002 \\
      VAR & \textcolor{red}{-0.000} & \textcolor{red}{-0.000} & \textcolor{red}{-0.001} & \textcolor{red}{-0.000} & 0.001 & \textcolor{red}{-0.001} & \textcolor{red}{-0.000} & \textcolor{red}{-0.000} & 0.001 & 0.000 & 0.000 & \textcolor{red}{-0.000} \\
      \hline
      Scenario 7 & & & & & & & & & & & & \\
      \hline
      ARIMA & 0.001 & 0.001 & 0.002 & 0.002 & 0.001 & 0.001 & 0.001 & 0.000 & 0.000 & \textcolor{red}{-0.001} & \textcolor{red}{-0.000} & \textcolor{red}{-0.000} \\
      ETS & 0.001 & 0.001 & 0.000 & 0.000 & 0.000 & 0.000 & 0.000 & 0.000 & 0.000 & 0.000 & 0.000 & 0.000 \\
      VAR & 0.000 & 0.000 & 0.000 & \textcolor{red}{-0.000} & 0.000 & \textcolor{red}{-0.000} & 0.001 & \textcolor{red}{-0.000} & 0.000 & \textcolor{red}{-0.000} & 0.000 & 0.000 \\
      \hline
      Scenario 8 & & & & & & & & & & & & \\
      \hline
      ARIMA & 0.001 & 0.000 & 0.000 & \textcolor{red}{-0.001} & 0.000 & 0.001 & 0.001 & 0.001 & 0.001 & 0.002 & 0.001 & 0.001 \\
      ETS & 0.000 & 0.000 & \textcolor{red}{-0.000} & 0.000 & \textcolor{red}{-0.000} & \textcolor{red}{-0.000} & \textcolor{red}{-0.000} & \textcolor{red}{-0.000} & \textcolor{red}{-0.000} & \textcolor{red}{-0.000} & \textcolor{red}{-0.000} & \textcolor{red}{-0.000} \\
      VAR & \textcolor{red}{-0.001} & \textcolor{red}{-0.002} & \textcolor{red}{-0.001} & \textcolor{red}{-0.003} & \textcolor{red}{-0.002} & \textcolor{red}{-0.003} & \textcolor{red}{-0.003} & \textcolor{red}{-0.004} & \textcolor{red}{-0.002} & \textcolor{red}{-0.003} & \textcolor{red}{-0.002} & \textcolor{red}{-0.003} \\
      \hline
      Scenario 9 & & & & & & & & & & & & \\
      \hline
      ARIMA & 0.001 & 0.002 & 0.001 & 0.001 & 0.001 & \textcolor{red}{-0.001} & 0.000 & \textcolor{red}{-0.000} & 0.000 & \textcolor{red}{-0.000} & \textcolor{red}{-0.000} & \textcolor{red}{-0.000} \\
      ETS & 0.001 & 0.002 & 0.002 & 0.002 & 0.002 & 0.001 & 0.001 & 0.001 & 0.001 & 0.001 & 0.001 & 0.001 \\
      VAR & \textcolor{red}{-0.001} & \textcolor{red}{-0.001} & \textcolor{red}{-0.001} & \textcolor{red}{-0.001} & \textcolor{red}{-0.001} & \textcolor{red}{-0.001} & \textcolor{red}{-0.000} & \textcolor{red}{-0.001} & \textcolor{red}{-0.000} & \textcolor{red}{-0.000} & \textcolor{red}{-0.000} & 0.000 \\
      \hline
    \end{tabular}%
  }
  \label{tab:sh_uni_mean}
\end{table}

%% file: Tabelas/Tabs_porc_sh_uni.tex
\begin{table}[!htb]
  \caption{Percentage of $\RelRMSE^{\Uni}$ greater than or equal to 0 considering the 16 series in the hierarchy and the 12 forecast horizons. Using the shrinkage approach to estimate $\bm{W}$ for the 9 scenarios.}
  \centering
  \begin{tabular}{lrrrrrrrrr}
    \hline
    \multirow{2}{*}{Model} & \multicolumn{9}{c}{Scenario} \\ \cline{2-10}
    & \multicolumn{1}{c}{1} & \multicolumn{1}{c}{2} & \multicolumn{1}{c}{3} & \multicolumn{1}{c}{4} & \multicolumn{1}{c}{5} & 6 & 7 & 8 & 9 \\ \hline
    ARIMA & 43.8 & 66.7 & 55.2 & 30.2 & 49.0 & 60.9 & 64.6 & 57.8 & 60.4 \\
    ETS & 87.0 & 40.1 & 93.8 & 56.2 & 55.2 & 92.2 & 78.6 & 40.6 & 90.1 \\
    VAR & 60.4 & 4.7 & 21.4 & 47.9 & 44.8 & 42.2 & 48.4 & 6.2 & 28.1 \\
    \hline
  \end{tabular}%
  \label{tab:porc_sh_uni}
\end{table}

%% file: Tabelas/Tabs_RMSSE.tex
\begin{table}[!htb]
\caption{Mean RMSSE of the reconciled multivariate forecasts by forecast horizon and base model for the nine analyzed scenarios, using the shrinkage approach to estimate $\boldsymbol{W}$. Bold values represent the lowest values per horizon and model in each scenario.}
\centering\resizebox{1\textwidth}{!}{%
\begin{tabular}{lcccccccccccc}\hline
\multirow{2}{*}{Model} & \multicolumn{12}{c}{Forecast horizons ($h$)}
\\ \cline{2-13}
 & 1 & 2 & 3 & 4 & 5 & 6 & 7 & 8 & 9 & 10 & 11 & 12 \\
\hline
Scenario 1\\
\hline
ARIMA & 0.635 & \textbf{0.797} & 0.883 & 0.944 & 1.000 & 1.030 & 1.055 & 1.069 & 1.113 & 1.123 & 1.147 & 1.141\\
ETS & \textbf{0.622} & 0.804 & 0.908 & 0.990 & 1.047 & 1.095 & 1.134 & 1.162 & 1.203 & 1.227 & 1.261 & 1.265\\
VAR & 0.649 & 0.816 & \textbf{0.862} & \textbf{0.901} & \textbf{0.979} & \textbf{1.017} & \textbf{1.018} & \textbf{1.012} & \textbf{1.067} & \textbf{1.068} & \textbf{1.084} & \textbf{1.054}\\
\hline
Scenario 2\\
\hline
ARIMA & 0.582 & 0.772 & \textbf{0.894} & 0.974 & \textbf{1.041} & \textbf{1.104} & 1.141 & 1.173 & 1.238 & 1.253 & 1.286 & 1.276\\
ETS & \textbf{0.564} & \textbf{0.769} & 0.903 & 1.002 & 1.077 & 1.154 & 1.209 & 1.259 & 1.321 & 1.353 & 1.389 & 1.392\\
VAR & 0.646 & 0.855 & 0.916 & \textbf{0.959} & 1.061 & 1.113 & \textbf{1.128} & \textbf{1.127} & \textbf{1.200} & \textbf{1.194} & \textbf{1.218} & \textbf{1.169}\\
\hline
Scenario 3\\
\hline
ARIMA & 0.675 & 0.793 & 0.846 & 0.860 & 0.881 & 0.890 & 0.894 & 0.901 & 0.919 & 0.904 & 0.929 & 0.922\\
ETS & 0.681 & 0.831 & 0.907 & 0.947 & 0.968 & 0.992 & 1.004 & 1.017 & 1.029 & 1.034 & 1.063 & 1.055\\
VAR & \textbf{0.659} & \textbf{0.774} & \textbf{0.806} & \textbf{0.817} & \textbf{0.842} & \textbf{0.859} & \textbf{0.853} & \textbf{0.853} & \textbf{0.868} & \textbf{0.859} & \textbf{0.876} & \textbf{0.860}\\
\hline
Scenario 4\\
\hline
ARIMA & 0.634 & \textbf{0.786} & 0.891 & 0.936 & \textbf{0.983} & 1.019 & 1.038 & 1.064 & 1.104 & 1.115 & 1.132 & 1.128\\
ETS & \textbf{0.617} & 0.789 & 0.909 & 0.984 & 1.031 & 1.085 & 1.120 & 1.162 & 1.203 & 1.231 & 1.255 & 1.258\\
VAR & 0.657 & 0.813 & \textbf{0.876} & \textbf{0.907} & 0.983 & \textbf{1.016} & \textbf{1.023} & \textbf{1.022} & \textbf{1.078} & \textbf{1.071} & \textbf{1.084} & \textbf{1.054}\\
\hline
Scenario 5\\
\hline
ARIMA & 0.587 & 0.766 & 0.895 & 0.979 & \textbf{1.056} & \textbf{1.113} & 1.143 & 1.186 & 1.235 & 1.262 & 1.279 & 1.285\\
ETS & \textbf{0.563} & \textbf{0.754} & \textbf{0.894} & 1.000 & 1.082 & 1.153 & 1.200 & 1.257 & 1.304 & 1.349 & 1.373 & 1.386\\
VAR & 0.642 & 0.842 & 0.912 & \textbf{0.958} & 1.085 & 1.124 & \textbf{1.138} & \textbf{1.128} & \textbf{1.212} & \textbf{1.203} & \textbf{1.219} & \textbf{1.181}\\
\hline
Scenario 6\\
\hline
ARIMA & 0.669 & 0.786 & 0.841 & 0.856 & 0.879 & 0.871 & 0.891 & 0.891 & 0.913 & 0.908 & 0.917 & 0.929\\
ETS & 0.675 & 0.813 & 0.893 & 0.931 & 0.964 & 0.969 & 0.992 & 0.999 & 1.029 & 1.036 & 1.048 & 1.054\\
VAR & \textbf{0.658} & \textbf{0.776} & \textbf{0.808} & \textbf{0.819} & \textbf{0.851} & \textbf{0.843} & \textbf{0.850} & \textbf{0.846} & \textbf{0.868} & \textbf{0.869} & \textbf{0.866} & \textbf{0.867}\\
\hline
Scenario 7\\
\hline
ARIMA & 0.639 & \textbf{0.796} & 0.892 & 0.937 & 0.996 & 1.036 & 1.056 & 1.072 & 1.106 & 1.120 & 1.145 & 1.147\\
ETS & \textbf{0.625} & 0.800 & 0.912 & 0.979 & 1.043 & 1.096 & 1.128 & 1.163 & 1.197 & 1.226 & 1.252 & 1.266\\
VAR & 0.656 & 0.815 & \textbf{0.866} & \textbf{0.889} & \textbf{0.972} & \textbf{1.019} & \textbf{1.013} & \textbf{1.005} & \textbf{1.049} & \textbf{1.063} & \textbf{1.075} & \textbf{1.055}\\
\hline
Scenario 8\\
\hline
ARIMA & 0.594 & 0.776 & \textbf{0.894} & 0.968 & \textbf{1.050} & \textbf{1.107} & 1.148 & 1.190 & 1.232 & 1.259 & 1.276 & 1.281\\
ETS & \textbf{0.573} & \textbf{0.767} & 0.903 & 0.995 & 1.080 & 1.150 & 1.206 & 1.264 & 1.308 & 1.349 & 1.376 & 1.390\\
VAR & 0.657 & 0.853 & 0.912 & \textbf{0.952} & 1.060 & 1.115 & \textbf{1.126} & \textbf{1.135} & \textbf{1.184} & \textbf{1.195} & \textbf{1.203} & \textbf{1.180}\\
\hline
Scenario 9\\
\hline
ARIMA & 0.683 & 0.790 & 0.844 & 0.853 & 0.886 & 0.898 & 0.904 & 0.902 & 0.918 & 0.917 & 0.921 & 0.926\\
ETS & 0.685 & 0.821 & 0.900 & 0.933 & 0.968 & 0.991 & 1.005 & 1.016 & 1.032 & 1.038 & 1.046 & 1.059\\
VAR & \textbf{0.666} & \textbf{0.770} & \textbf{0.805} & \textbf{0.806} & \textbf{0.843} & \textbf{0.859} & \textbf{0.853} & \textbf{0.851} & \textbf{0.861} & \textbf{0.866} & \textbf{0.864} & \textbf{0.869}\\
\hline\end{tabular}}\label{tab:RMSSE}\end{table}

%% file: Tabelas/Tabs_adm_sh.tex
\begin{table}[!htb]
  \caption{$\RelRMSE^{\Base}$ of admission series. ARIMA model for base forecasts and the shrinkage approach to estimate $\bm{W}$. Values in red indicate a $\RelRMSE^{\Base}$ less than 0.}
  \centering
  \resizebox{1\textwidth}{!}{%
    \begin{tabular}{lrrrrrrrrrrrr}
      \hline
      \multirow{2}{*}{Series} & \multicolumn{12}{c}{Forecast horizons ($h$)} \\ \cline{2-13}
      & 1 & 2 & 3 & 4 & 5 & 6 & 7 & 8 & 9 & 10 & 11 & 12 \\ \hline
      Total & 0.020 & 0.168 & 0.162 & 0.206 & 0.235 & 0.245 & 0.278 & 0.282 & 0.360 & 0.326 & 0.362 & 0.281 \\
      Midwest & \textcolor{red}{-0.001} & 0.101 & 0.106 & 0.179 & 0.255 & 0.299 & 0.340 & 0.371 & 0.465 & 0.505 & 0.569 & 0.619 \\
      Northeast & 0.087 & 0.054 & 0.016 & \textcolor{red}{-0.064} & \textcolor{red}{-0.120} & \textcolor{red}{-0.161} & \textcolor{red}{-0.038} & 0.026 & 0.236 & 0.200 & 0.242 & 0.143 \\
      North & 0.122 & 0.057 & 0.078 & 0.134 & 0.180 & 0.254 & 0.358 & 0.311 & 0.340 & 0.275 & 0.341 & 0.356 \\
      Southeast & 0.032 & 0.139 & 0.079 & 0.089 & 0.150 & 0.168 & 0.224 & 0.245 & 0.329 & 0.279 & 0.222 & 0.183 \\
      South & \textcolor{red}{-0.290} & \textcolor{red}{-0.169} & \textcolor{red}{-0.074} & \textcolor{red}{-0.090} & \textcolor{red}{-0.174} & \textcolor{red}{-0.141} & \textcolor{red}{-0.136} & \textcolor{red}{-0.103} & \textcolor{red}{-0.086} & \textcolor{red}{-0.012} & 0.150 & 0.156 \\
      AC & 0.036 & 0.049 & 0.049 & 0.072 & 0.104 & 0.108 & 0.142 & 0.140 & 0.145 & 0.156 & 0.133 & 0.172 \\
      AL & 0.021 & 0.012 & 0.014 & 0.054 & 0.103 & 0.102 & 0.089 & 0.069 & 0.020 & 0.061 & 0.036 & \textcolor{red}{-0.021} \\
      AM & \textcolor{red}{-0.062} & \textcolor{red}{-0.008} & 0.026 & 0.090 & 0.143 & 0.189 & 0.268 & 0.216 & 0.362 & 0.228 & 0.364 & 0.376 \\
      AP & 0.004 & 0.023 & 0.059 & 0.073 & 0.103 & 0.105 & 0.116 & 0.098 & 0.105 & 0.117 & 0.079 & 0.056 \\
      BA & \textcolor{red}{-0.002} & 0.077 & 0.072 & 0.119 & 0.218 & 0.279 & 0.318 & 0.320 & 0.336 & 0.393 & 0.412 & 0.490 \\
      CE & \textcolor{red}{-0.016} & 0.090 & 0.081 & 0.120 & 0.143 & 0.163 & 0.197 & 0.200 & 0.218 & 0.188 & 0.116 & \textcolor{red}{-0.033} \\
      DF & 0.008 & 0.045 & 0.060 & 0.112 & 0.171 & 0.199 & 0.202 & 0.182 & 0.229 & 0.214 & 0.302 & 0.400 \\
      ES & \textcolor{red}{-0.028} & 0.021 & 0.035 & 0.071 & 0.057 & \textcolor{red}{-0.004} & \textcolor{red}{-0.054} & \textcolor{red}{-0.254} & \textcolor{red}{-0.190} & \textcolor{red}{-0.206} & \textcolor{red}{-0.051} & \textcolor{red}{-0.045} \\
      GO & 0.030 & 0.082 & 0.079 & 0.101 & 0.148 & 0.178 & 0.233 & 0.249 & 0.322 & 0.346 & 0.395 & 0.428 \\
      MA & 0.028 & 0.076 & 0.071 & 0.109 & 0.145 & 0.166 & 0.203 & 0.260 & 0.343 & 0.336 & 0.358 & 0.368 \\
      MG & 0.058 & 0.011 & 0.001 & 0.002 & 0.010 & 0.008 & 0.014 & 0.033 & 0.058 & 0.083 & 0.108 & 0.108 \\
      MS & 0.000 & 0.007 & 0.023 & 0.021 & 0.047 & 0.022 & 0.021 & 0.056 & 0.076 & 0.110 & 0.090 & 0.093 \\
      MT & 0.048 & 0.072 & 0.106 & 0.121 & 0.160 & 0.162 & 0.199 & 0.223 & 0.243 & 0.238 & 0.252 & 0.306 \\
      PA & 0.001 & 0.008 & \textcolor{red}{-0.030} & \textcolor{red}{-0.007} & \textcolor{red}{-0.035} & \textcolor{red}{-0.073} & \textcolor{red}{-0.081} & \textcolor{red}{-0.126} & \textcolor{red}{-0.128} & \textcolor{red}{-0.170} & \textcolor{red}{-0.267} & \textcolor{red}{-0.429} \\
      PB & \textcolor{red}{-0.067} & 0.010 & \textcolor{red}{-0.002} & 0.057 & 0.083 & 0.110 & 0.095 & \textcolor{red}{-0.004} & \textcolor{red}{-0.153} & \textcolor{red}{-0.208} & \textcolor{red}{-0.410} & \textcolor{red}{-0.420} \\
      PE & \textcolor{red}{-0.015} & \textcolor{red}{-0.035} & \textcolor{red}{-0.037} & \textcolor{red}{-0.027} & \textcolor{red}{-0.030} & \textcolor{red}{-0.043} & \textcolor{red}{-0.083} & \textcolor{red}{-0.131} & \textcolor{red}{-0.243} & \textcolor{red}{-0.258} & \textcolor{red}{-0.246} & \textcolor{red}{-0.299} \\
      PI & 0.049 & 0.111 & 0.079 & 0.101 & 0.119 & 0.139 & 0.137 & 0.110 & 0.053 & \textcolor{red}{-0.133} & \textcolor{red}{-0.491} & \textcolor{red}{-1.059} \\
      PR & 0.050 & 0.041 & 0.007 & \textcolor{red}{-0.006} & 0.071 & 0.142 & 0.260 & 0.303 & 0.329 & 0.296 & 0.289 & 0.241 \\
      RJ & \textcolor{red}{-0.015} & 0.093 & 0.025 & 0.108 & 0.213 & 0.242 & 0.280 & 0.260 & 0.438 & 0.279 & 0.386 & 0.231 \\
      RN & \textcolor{red}{-0.027} & 0.024 & \textcolor{red}{-0.007} & \textcolor{red}{-0.021} & 0.007 & \textcolor{red}{-0.020} & \textcolor{red}{-0.065} & \textcolor{red}{-0.148} & \textcolor{red}{-0.365} & \textcolor{red}{-0.484} & \textcolor{red}{-0.895} & \textcolor{red}{-1.114} \\
      RO & \textcolor{red}{-0.003} & 0.040 & 0.076 & 0.135 & 0.200 & 0.206 & 0.224 & 0.243 & 0.298 & 0.352 & 0.459 & 0.555 \\
      RR & \textcolor{red}{-0.007} & 0.014 & 0.002 & \textcolor{red}{-0.006} & 0.002 & \textcolor{red}{-0.013} & 0.010 & 0.013 & 0.045 & 0.030 & 0.054 & 0.065 \\
      RS & 0.059 & 0.020 & 0.029 & 0.031 & 0.035 & 0.023 & 0.026 & 0.016 & \textcolor{red}{-0.011} & \textcolor{red}{-0.010} & \textcolor{red}{-0.027} & \textcolor{red}{-0.085} \\
      SC & 0.082 & 0.123 & 0.144 & 0.140 & 0.173 & 0.167 & 0.206 & 0.168 & 0.140 & 0.116 & 0.051 & \textcolor{red}{-0.112} \\
      SE & \textcolor{red}{-0.013} & 0.003 & 0.005 & 0.027 & 0.029 & 0.083 & 0.036 & 0.003 & \textcolor{red}{-0.075} & \textcolor{red}{-0.202} & \textcolor{red}{-0.426} & \textcolor{red}{-0.806} \\
      SP & \textcolor{red}{-0.088} & \textcolor{red}{-0.024} & 0.006 & 0.036 & \textcolor{red}{-0.007} & \textcolor{red}{-0.026} & \textcolor{red}{-0.064} & \textcolor{red}{-0.069} & \textcolor{red}{-0.265} & \textcolor{red}{-0.091} & \textcolor{red}{-0.014} & \textcolor{red}{-0.015} \\
      TO & 0.021 & 0.045 & 0.046 & 0.078 & 0.114 & 0.130 & 0.163 & 0.166 & 0.156 & 0.104 & 0.008 & \textcolor{red}{-0.084} \\
      \hline
    \end{tabular}%
  }
  \label{tab:adm_sh}
\end{table}

%% file: Tabelas/Tabs_dem_sh.tex
\begin{table}[!htb]
  \caption{$\RelRMSE^{\Base}$ of dismissal series. ARIMA model for base forecasts and the shrinkage approach to estimate $\bm{W}$. Values in red indicate a $\RelRMSE^{\Base}$ less than 0.}
  \centering
  \resizebox{1\textwidth}{!}{%
    \begin{tabular}{lrrrrrrrrrrrr}
      \hline
      \multirow{2}{*}{Series} & \multicolumn{12}{c}{Forecast horizons ($h$)} \\ \cline{2-13}
      & 1 & 2 & 3 & 4 & 5 & 6 & 7 & 8 & 9 & 10 & 11 & 12 \\ \hline
      Total & \textcolor{red}{-0.053} & \textcolor{red}{-0.008} & 0.029 & 0.058 & 0.018 & 0.030 & 0.012 & \textcolor{red}{-0.016} & \textcolor{red}{-0.005} & 0.063 & 0.087 & 0.046 \\
      Midwest & \textcolor{red}{-0.006} & 0.006 & 0.024 & 0.083 & 0.124 & 0.110 & 0.095 & 0.156 & 0.205 & 0.221 & 0.206 & 0.189 \\
      Northeast & \textcolor{red}{-0.050} & \textcolor{red}{-0.189} & 0.045 & \textcolor{red}{-0.048} & \textcolor{red}{-0.150} & \textcolor{red}{-0.130} & \textcolor{red}{-0.146} & \textcolor{red}{-0.147} & \textcolor{red}{-0.405} & \textcolor{red}{-0.402} & \textcolor{red}{-0.187} & \textcolor{red}{-0.512} \\
      North & \textcolor{red}{-0.052} & 0.032 & \textcolor{red}{-0.005} & \textcolor{red}{-0.078} & \textcolor{red}{-0.127} & \textcolor{red}{-0.102} & \textcolor{red}{-0.046} & \textcolor{red}{-0.055} & \textcolor{red}{-0.050} & \textcolor{red}{-0.024} & \textcolor{red}{-0.019} & \textcolor{red}{-0.048} \\
      Southeast & \textcolor{red}{-0.090} & \textcolor{red}{-0.056} & 0.032 & 0.035 & \textcolor{red}{-0.001} & \textcolor{red}{-0.008} & \textcolor{red}{-0.049} & \textcolor{red}{-0.064} & \textcolor{red}{-0.091} & \textcolor{red}{-0.052} & \textcolor{red}{-0.043} & \textcolor{red}{-0.036} \\
      South & \textcolor{red}{-0.024} & \textcolor{red}{-0.067} & 0.037 & 0.013 & 0.062 & 0.045 & 0.019 & 0.040 & 0.038 & 0.157 & 0.196 & 0.299 \\
      AC & \textcolor{red}{-0.016} & 0.013 & 0.013 & 0.030 & 0.045 & 0.030 & \textcolor{red}{-0.008} & \textcolor{red}{-0.009} & \textcolor{red}{-0.033} & \textcolor{red}{-0.008} & 0.036 & 0.028 \\
      AL & 0.033 & 0.050 & 0.104 & 0.036 & \textcolor{red}{-0.013} & 0.025 & 0.059 & 0.108 & 0.181 & 0.162 & 0.132 & 0.182 \\
      AM & \textcolor{red}{-0.036} & \textcolor{red}{-0.016} & \textcolor{red}{-0.012} & \textcolor{red}{-0.104} & \textcolor{red}{-0.164} & \textcolor{red}{-0.149} & \textcolor{red}{-0.139} & \textcolor{red}{-0.165} & \textcolor{red}{-0.236} & \textcolor{red}{-0.347} & \textcolor{red}{-0.169} & \textcolor{red}{-0.148} \\
      AP & 0.007 & 0.017 & 0.022 & 0.003 & 0.007 & \textcolor{red}{-0.003} & 0.019 & 0.012 & 0.024 & 0.032 & 0.042 & 0.068 \\
      BA & 0.016 & 0.010 & \textcolor{red}{-0.029} & \textcolor{red}{-0.045} & \textcolor{red}{-0.040} & \textcolor{red}{-0.011} & 0.034 & \textcolor{red}{-0.022} & 0.059 & 0.089 & 0.074 & 0.063 \\
      CE & \textcolor{red}{-0.002} & \textcolor{red}{-0.001} & \textcolor{red}{-0.003} & \textcolor{red}{-0.028} & 0.060 & 0.001 & \textcolor{red}{-0.006} & \textcolor{red}{-0.109} & \textcolor{red}{-0.122} & \textcolor{red}{-0.126} & \textcolor{red}{-0.161} & \textcolor{red}{-0.008} \\
      DF & \textcolor{red}{-0.046} & \textcolor{red}{-0.043} & 0.003 & 0.028 & 0.043 & \textcolor{red}{-0.001} & \textcolor{red}{-0.014} & \textcolor{red}{-0.046} & \textcolor{red}{-0.044} & \textcolor{red}{-0.003} & 0.030 & 0.059 \\
      ES & 0.017 & 0.049 & 0.077 & 0.092 & 0.118 & 0.073 & 0.029 & \textcolor{red}{-0.013} & \textcolor{red}{-0.017} & \textcolor{red}{-0.003} & \textcolor{red}{-0.034} & \textcolor{red}{-0.102} \\
      GO & \textcolor{red}{-0.012} & \textcolor{red}{-0.021} & 0.103 & 0.078 & 0.086 & 0.043 & \textcolor{red}{-0.001} & \textcolor{red}{-0.014} & \textcolor{red}{-0.040} & \textcolor{red}{-0.032} & \textcolor{red}{-0.021} & \textcolor{red}{-0.021} \\
      MA & \textcolor{red}{-0.056} & \textcolor{red}{-0.022} & 0.075 & \textcolor{red}{-0.011} & \textcolor{red}{-0.096} & \textcolor{red}{-0.046} & 0.022 & \textcolor{red}{-0.092} & \textcolor{red}{-0.082} & \textcolor{red}{-0.045} & \textcolor{red}{-0.024} & \textcolor{red}{-0.159} \\
      MG & \textcolor{red}{-0.131} & \textcolor{red}{-0.061} & 0.029 & 0.072 & 0.114 & 0.115 & 0.085 & 0.069 & 0.037 & 0.071 & 0.049 & \textcolor{red}{-0.015} \\
      MS & 0.019 & \textcolor{red}{-0.001} & 0.035 & 0.059 & 0.073 & 0.046 & 0.052 & 0.041 & 0.035 & 0.098 & 0.102 & 0.124 \\
      MT & 0.023 & 0.056 & 0.139 & 0.193 & 0.262 & 0.251 & 0.275 & 0.382 & 0.390 & 0.407 & 0.419 & 0.464 \\
      PA & \textcolor{red}{-0.013} & \textcolor{red}{-0.008} & \textcolor{red}{-0.004} & \textcolor{red}{-0.023} & \textcolor{red}{-0.024} & \textcolor{red}{-0.000} & 0.032 & 0.005 & 0.010 & 0.011 & 0.049 & \textcolor{red}{-0.028} \\
      PB & 0.063 & 0.068 & 0.061 & 0.066 & 0.081 & 0.030 & 0.061 & 0.097 & 0.041 & 0.090 & 0.108 & 0.108 \\
      PE & 0.033 & 0.053 & 0.073 & 0.071 & 0.054 & \textcolor{red}{-0.012} & \textcolor{red}{-0.030} & \textcolor{red}{-0.175} & \textcolor{red}{-0.177} & \textcolor{red}{-0.171} & \textcolor{red}{-0.148} & \textcolor{red}{-0.110} \\
      PI & \textcolor{red}{-0.042} & \textcolor{red}{-0.031} & 0.014 & \textcolor{red}{-0.061} & \textcolor{red}{-0.069} & \textcolor{red}{-0.047} & 0.035 & \textcolor{red}{-0.052} & \textcolor{red}{-0.022} & 0.018 & 0.048 & \textcolor{red}{-0.115} \\
      PR & \textcolor{red}{-0.027} & \textcolor{red}{-0.007} & 0.028 & 0.047 & 0.049 & 0.019 & 0.017 & 0.003 & \textcolor{red}{-0.003} & 0.011 & 0.021 & \textcolor{red}{-0.014} \\
      RJ & \textcolor{red}{-0.165} & \textcolor{red}{-0.099} & \textcolor{red}{-0.002} & 0.091 & 0.077 & 0.056 & 0.057 & 0.045 & 0.025 & 0.101 & 0.120 & 0.093 \\
      RN & \textcolor{red}{-0.087} & \textcolor{red}{-0.068} & 0.028 & 0.009 & \textcolor{red}{-0.026} & \textcolor{red}{-0.034} & 0.058 & 0.075 & 0.086 & 0.204 & 0.226 & 0.172 \\
      RO & \textcolor{red}{-0.026} & \textcolor{red}{-0.010} & \textcolor{red}{-0.010} & 0.002 & 0.054 & 0.032 & 0.005 & \textcolor{red}{-0.024} & \textcolor{red}{-0.051} & \textcolor{red}{-0.096} & \textcolor{red}{-0.129} & \textcolor{red}{-0.210} \\
      RR & \textcolor{red}{-0.059} & \textcolor{red}{-0.047} & \textcolor{red}{-0.010} & \textcolor{red}{-0.030} & \textcolor{red}{-0.026} & \textcolor{red}{-0.031} & \textcolor{red}{-0.019} & 0.017 & 0.067 & 0.084 & 0.076 & 0.062 \\
      RS & \textcolor{red}{-0.017} & \textcolor{red}{-0.016} & 0.018 & 0.039 & 0.020 & 0.007 & 0.008 & \textcolor{red}{-0.003} & \textcolor{red}{-0.010} & 0.051 & 0.090 & 0.086 \\
      SC & 0.034 & 0.023 & 0.046 & 0.039 & 0.038 & 0.025 & 0.020 & \textcolor{red}{-0.002} & 0.007 & 0.063 & 0.087 & 0.063 \\
      SE & 0.001 & \textcolor{red}{-0.019} & \textcolor{red}{-0.020} & \textcolor{red}{-0.053} & \textcolor{red}{-0.049} & \textcolor{red}{-0.055} & \textcolor{red}{-0.044} & 0.021 & 0.042 & \textcolor{red}{-0.066} & 0.006 & 0.066 \\
      SP & \textcolor{red}{-0.113} & \textcolor{red}{-0.019} & 0.047 & 0.010 & \textcolor{red}{-0.014} & \textcolor{red}{-0.024} & \textcolor{red}{-0.055} & \textcolor{red}{-0.060} & \textcolor{red}{-0.068} & \textcolor{red}{-0.035} & \textcolor{red}{-0.022} & \textcolor{red}{-0.029} \\
      TO & 0.003 & \textcolor{red}{-0.022} & 0.080 & 0.013 & 0.024 & 0.007 & 0.014 & \textcolor{red}{-0.036} & \textcolor{red}{-0.020} & \textcolor{red}{-0.031} & \textcolor{red}{-0.033} & \textcolor{red}{-0.078} \\
      \hline
    \end{tabular}%
  }
  \label{tab:dem_sh}
\end{table}